\documentclass{aa}  
\usepackage{graphicx}
\usepackage{txfonts}
\PassOptionsToPackage{hyphens}{url}
\usepackage[breaklinks]{hyperref}
\usepackage{xurl}
\usepackage{changes}
\usepackage{natbib}
\usepackage{pdflscape}
\usepackage{longtable}
\usepackage{supertabular,booktabs}
\usepackage{multicol}
\usepackage{todonotes}
\usepackage{longtable}
\usepackage{orcidlink} 

\def\teff{$T\rm_{eff}$}

\begin{document} 

\title{Two new Li-rich dwarf stars in M\,4: \\ abundance patterns and possible origins.}
\titlerunning{}

\author{F.~Lucertini\inst{\ref{eso}}\orcidlink{0000-0002-8785-0109}
    \and
    E.~Pancino\inst{\ref{oaa}}\orcidlink{0000-0003-0788-5879}
    \and
    L.~Monaco\inst{\ref{bello}}\orcidlink{0000-0002-3148-9836}
    \and
    L.~Sbordone\inst{\ref{eso}}\orcidlink{0000-0002-2285-8708}
    \and
    I.~Baeza\inst{\ref{eso},\ref{bello}}\orcidlink{0000-0002-9881-6336}
    \and
    D.~A.~Alvarez Garay\inst{\ref{oaa},}\orcidlink{0000-0002-9156-0729}
    \and
    S.~Villanova\inst{\ref{bello}}\orcidlink{0000-0001-6205-1493}
    \and
    A.~Mucciarelli\inst{\ref{unibo}}\orcidlink{0000-0001-9158-8580}
    \and
    N.~Sanna\inst{\ref{oaa}}\orcidlink{0000-0001-9275-9492}
}

\institute{
    ESO - European Southern Observatory, Alonso de Cordova 3107, Vitacura, Santiago, Chile\label{eso}
    \and
    INAF -- Osservatorio Astrofisico di Arcetri, I-52025, Firenze, Italy\label{oaa}
    \and
    Universidad Andres Bello, Facultad de Ciencias Exactas, Departamento de F{\'i}sica y Astronom{\'i}a - Instituto de Astrofisica, Autopista Concepci\'on-Talcahuano, 7100, Talcahuano, Chile\label{bello}
    \and
    Dipartimento di Fisica e Astronomia “Augusto Righi”, Alma Mater Studiorum, Università di Bologna, Via Gobetti 93/2, I-40129
    Bologna, Italy\label{unibo}
}

   \date{Received 19 October 2026 / Accepted 14 November 2026}

  \abstract
{Li-rich stars are rare, and only 16 are known in globular clusters (GCs), spanning evolutionary stages from the main sequence turnoff to the red giant branch. Their origin is still debated, but GCs provide a unique environment to discriminate among proposed mechanisms for Li enrichment.}
{We aim to characterise two newly discovered Li-rich dwarf stars, 44606 and 58195, in the GC M 4 (NGC 6121) and investigate their origin among the formation channels proposed in the literature. The previously known Li-rich star in the cluster, namely 37934, was re-analyzed for comparison.}
{We performed a standard abundance analysis based on 1D model atmospheres in local thermodynamic equilibrium (LTE) on high-resolution Ultraviolet and Visual Echelle Spectrograph (UVES). We measured the projected rotational velocity and searched for chromospheric activity.}
{We derive A(Li)= 2.58, 2.51\,dex for stars 44606 and 58195, respectively, significantly different from the A(Li) distribution of normal dwarfs in M\,4. For the known Li-rich star 37934, we obtain A(Li) = 3.01\,dex, consistent with previous work. We estimate detailed abundances of 24 species (Li, C, N, O, Na, Mg, Al, Si, S, K, Ca, Sc, Ti, V, Cr, Mn, Fe, Co, Ni, Sr, Y, Zr, Ba, and Nd) in the three stars. None of our targets show detectable rotational broadening or chromospheric activity.}
{The Li-rich stars analysed here are unevolved, ruling out in-situ Li production. Planet engulfment is very unlikely in GCs, and we detect no present-day rotation or activity. By exclusion, the most plausible remaining scenarios involve different types of interaction with a companion, including dynamical interactions, mass transfer, and mergers. Currently, we have not detected a companion.}

\keywords{stars: abundances -- (Galaxy:) globular clusters: individual:: M 4 -- techniques: spectroscopic}
   \maketitle
\section{Introduction}
\label{sec:intro}

Lithium (Li) is one of the few species produced during Big Bang nucleosynthesis, which predicts a cosmological abundance of A(Li)\,$\simeq$\,2.7\,dex \citep{2011ARNPS..61...47F, Fields_2020}. 
However, the typical Li abundance encountered even in the least processed stellar environments, i.e., metal-poor dwarf stars in the halo, is significantly lower, A(Li)\,$\simeq$\,2.2\,dex \citep{1982A&A...115..357S}. 
One possible explanation is atomic diffusion, where gravitational settling transports Li from the surface layers into the stellar interior, leading to a lower observed surface abundance \citep{2006Natur.442..657K}.
After the main sequence, the surface Li of low-mass stars is further diluted by two mixing episodes: the first dredge-up at the base of the red giant branch (RGB) and the RGB bump \citep{2000A&A...354..169G,2009A&A...503..545L,2012MNRAS.419.2195M}.

Within this framework, we need to better understand the mechanisms that alter surface Li abundance in stars. In this respect, Li-rich stars, i.e., stars significantly richer in Li than similar stars, offer a unique perspective. 
The frequency of Li-rich stars is $\lesssim$\,1\% of the disk stars in the Galactic field population and in open clusters  \citep{2020MNRAS.494.1348D} and most of them are in the red clump evolutionary phase, i.e., they are burning helium in the core. Different explanations have been proposed for the existence of Li-rich stars, including: Li production through the Cameron-Fowler or other related mechanisms \citep{1996ApJ...456L.115D,2010A&A...522A..10C,2010ApJ...719...28D}; mass transfer from a companion able to produce Li, or Li enhancement caused by dynamical interaction with binary companions \citep{2019ApJ...880..125C}; engulfment of a planet or other sub-stellar companion \citep{1999MNRAS.308.1133S}; or production by novae \citep{2021A&A...653A..72R,2025A&A...699A.171M}. 
Another possibility is that Li, mostly $^6$Li, is produced in flares in the atmosphere of stars \citep{2000ApJ...534L.207R,2007A&A...469..265T,2017A&A...604A..44M}, although it is difficult to explain the entirety of the Li observed in Li-rich stars this way. 
No firm conclusion has been reached on whether a single mechanism explains all cases, or whether different mechanisms operate at different evolutionary stages or simultaneously.

\begin{figure*}[t]
    \centering
    \includegraphics[width=0.95\textwidth]{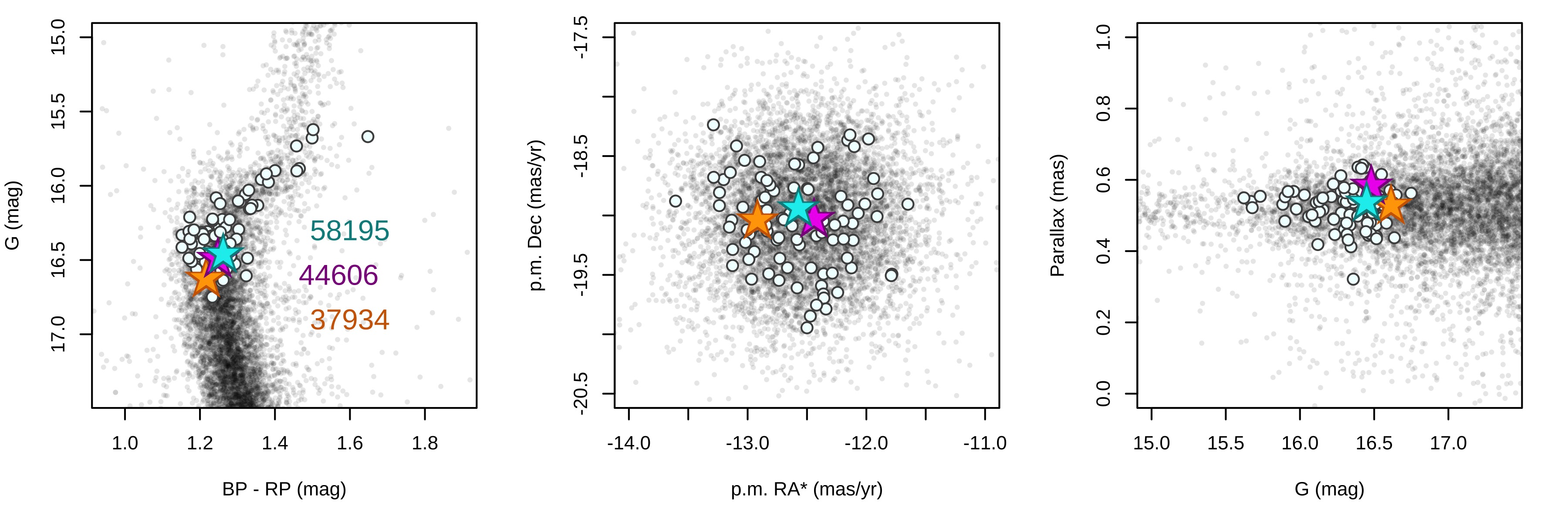}
    \caption{Properties of target stars, indicated as large colored symbols in each panel. The {\em Gaia} DR3 member stars in M\,4 \citep{2021MNRAS.505.5978V} are plotted in the background as small gray symbols. White circles represent the sample published by \citet{2012A&A...539A.157M}. }
    \label{fig:targets}
\end{figure*}

\begin{table*}[h]
    \centering
    \caption{Observed sample with basic properties.}
    \begin{tabular}{llccccccl}
    \hline \hline
         Gaia ID & Star ID & RA & Dec & G & BP--RP & RV & $\sigma_{\rm RV}$ & S/N \\
         &&[hh:mm:ss]&[$\deg$:mm:ss]&[mag]&[mag]& [km s$^{-1}$] & [km s$^{-1}$] & [@800\,nm] \\
         \hline
        6045462826174755328 & 37934 & 16:23:23.62 & --26:37:41.5 & 16.61 & 1.22 & 66.9 & 0.9 & 34\\
        6045465162637456128 & 44606 & 16:23:54.10 & --26:32:52.4 & 16.48 & 1.25 & 75.3 & 0.4 & 31 \\
        6045478494216476928 & 58195 & 16:23:38.34 & --26:26:19.0 & 16.45 & 1.26 & 68.6 & 0.3 & 27\\
         \hline \hline
    \end{tabular} 
    \label{tab:obs}
\end{table*}

Globular clusters (GCs) represent a peculiar environment, a crowded ensemble of old, metal-poor, and low-mass stars that has been extensively explored spectroscopically, yielding so far only 16 Li-rich metal-poor stars (one in M\,5 \citealt{1998AJ....116.2984C}; two in NGC\,362 \citealt{1999ApJ...516L..73S, 2015ApJ...801L..32D}; one in M\,3 \citealt{1999ApJ...518L..53K}; one in M\,68 \citealt{2011ApJ...743..107R}; one in NGC\,6397 \citealt{2011ApJ...738L..29K}; one in M\,4 \citealt{2012A&A...539A.157M}; two in M\,68, one in NGC\,5053, one in NGC\,5897, two in M\,30 \citealt{2016ApJ...819..135K}; two in $\omega$Cen \citealt{2019A&A...623A..55M}, one in NGC\,1261 \citealt{2020A&A...639L...2S}).
Despite their much lower frequency than Li-rich stars in the field and in open clusters (0.2\% for the RGB, 1.6\% for the AGB, and 0.3\% for all giants; \citealt{2016ApJ...819..135K}), these Li-rich stars do not cluster around a single evolutionary stage. They span the full magnitude range, from the tip of the RGB to the main-sequence turnoff. Some additional considerations can be made in the case of GCs. First, planet engulfment seems unlikely as an explanation, since only one planet has been found so far in GCs \citep{2019MNRAS.489.4311C}, out of a present sample of about 4000 known exoplanets. Second, the contribution from novae through galactic chemical evolution can be excluded, since it starts to become important only above [Fe/H]\,$\simeq$\,--1\,dex \citep{2021A&A...653A..72R}, although we cannot exclude the pollution by a companion in a nova-like binary system \citep{2025A&A...699A.171M}. Third, the lack of concentration around a specific evolutionary stage makes direct Li production in the stars themselves less viable. This leaves binary interactions as the most likely explanation, either through mass transfer from a companion that can produce Li or through dynamical interactions with a companion star that can enhance surface Li abundance. This, of course, does not exclude the possibility that a few giants among the 16 known Li-rich stars actually produced Li themselves, but this should be rare and limited to red giants.

Additionally, GCs are also affected by the multiple populations (MP) phenomenon \citep{2015MNRAS.454.4197R,2018ARA&A..56...83B}. From the chemistry point of view, this manifests itself as a series of intrinsic spreads of elements, sometimes structured in anti-correlations of pairs of species \citep[C-N, Na-O, and Mg-Al, ][]{2004AJ....127.1579B,2009A&A...505..117C,2017A&A...601A.112P}) that are involved in the hot hydrogen burning chains associated with the CNO cycle \citep{1989ATsir1538...11D}. Therefore, we expect any chemical fingerprint of putative mechanisms producing Li-rich stars to overlap with MP chemistry, complicating interpretation.

In this paper, we present the discovery of two moderately Li-rich dwarfs in M\,4 (NGC\,6121), a GC already known to host one Li-rich star \citep{2012A&A...539A.157M}, bringing the sample of Li-rich stars in GCs to 18. We also re-analyze the Li-rich star previously discovered using medium-resolution FLAMES spectra \citep{2002Msngr.110....1P}, profiting from new UVES observations \citep{2000SPIE.4008..534D}. 
The paper is organized as follows: in Sect.~\ref{sec:data} we present the dataset, data preparation, and analysis; in Sect.~\ref{sec:res} we present the abundance analysis and we validate the results; in Sect.~\ref{sec:disc} we discuss how our results can be interpreted in the framework of the possible formation channels for Li-rich stars; and in Sect.~\ref{sec:concl} we summarize our results and draw our conclusions.

\section{Data and analysis}
\label{sec:data}

This work is based on high-resolution spectra of two newly discovered Li-rich dwarfs. 
In July 2012, we observed more than 400 main-sequence turn-off M\,4 stars in four plate configurations with the FLAMES/GIRAFFE setup LR6 (ESO program ID 089.D-0062(A), R=8500, wavelength range 643.8-718.4\,nm). We acquired three 2775\,s exposures for each plate configuration. After visually inspecting the acquired spectra, we selected stars 44606 and 58195 for high-resolution spectroscopic follow-up (ESO program ID 089.D-0062(B)), using UVES in stand-alone mode. The previously known Li-rich main-sequence star \citep[ID 37934,][]{2012A&A...539A.157M,2016A&A...594A..79S} was also observed as part of this program, as well as one additional target whose spectrum resulted, however, of low quality and was not further investigated.
The UVES data were obtained with a slit width of 1", resulting in a spectral resolution of R $\simeq$\,42\,310.
The signal-to-noise (S/N) ratio from the three spectra at about 800\,nm is around 30 (see Table~\ref{tab:obs}). The instrument was set with Dichroic$\#$2, with blue and red arms centered at 390 and 760 nm, respectively. The final dataset covers the wavelength range $\simeq$328 to 946\,nm, with gaps.

\begin{figure*}[t]
    \centering
    \includegraphics[width=0.9\textwidth]{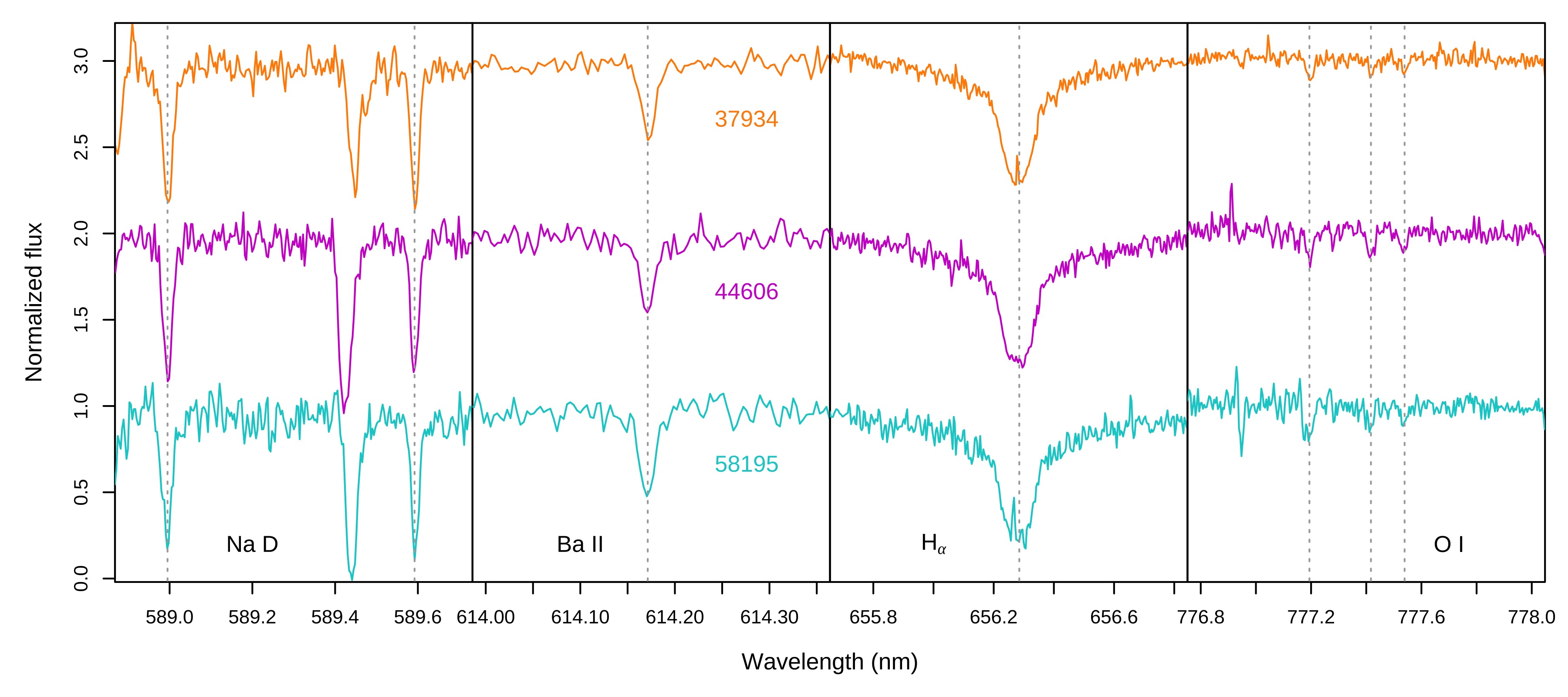}
    \caption{Some spectral regions of interest for the three observed stars, as annotated. In each panel, the spectra are offset for clarity. We plot the rest-frame, telluric-corrected, and normalized spectra. The vertical dashed lines represent the rest wavelengths of the indicated lines.}
    \label{fig:spec}
\end{figure*}

\subsection{Target characterization}
\label{sec:gaia}

The properties of the three stars considered here are illustrated in Fig.~\ref{fig:targets}, where it can be appreciated that they follow the typical behavior of M\,4 member stars in the color-magnitude diagram (CMD) and in the astrometric properties obtained from {\em Gaia} Data Release 3 (DR3, \citealt{gaia,gdr3}).

\begin{figure}[t]
    \centering
    \includegraphics[width=0.95\columnwidth]{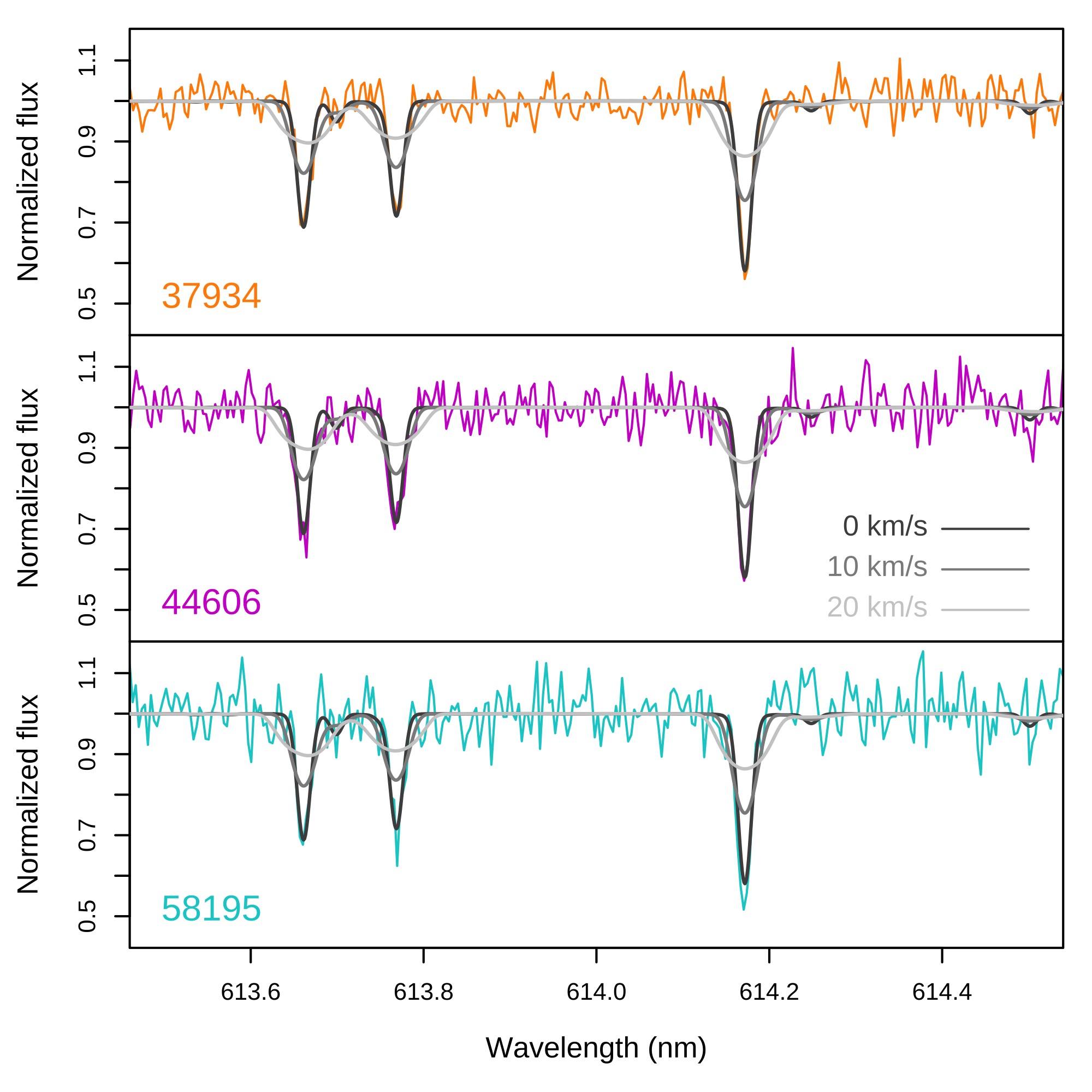}
    \caption{Example of rotationally broadened model fits. We used ATLAS\,9 models with the stellar parameters from Table~\ref{tab:pars} to generate synthetic spectra with three distinct broadenings of zero, 10, and 20\,km\,s$^{-1}$, as labelled. Each panel shows a different target.}
    \label{fig:rot}
\end{figure}

We searched for the {\tt ruwe}, {\tt ipd\_frac\_multi\_peak}, and {\tt ipd\_frac\_odd\_win} parameters in the {\em Gaia} DR3 catalog \footnote{We did not use the C* parameter \citep{2021A&A...649A...3R}, which is based on BP/RP spectrophotometry, because we are more interested in the contamination by sources within about 1", which is the width of the UVES slit employed, and because we only used Gaia data from the astrometric field in our analysis.}. A {\tt ruwe}\,$\gtrsim$\,1.4 indicates problems in the astrometric solution, possibly due to a companion star, among other effects \citep{2021A&A...649A...2L}. Our three targets have {\tt ruwe}\,=\, 1.02, 1.04, and 1.04.
The parameters {\tt ipd\_frac\_multi\_peak} and {\tt ipd\_frac\_odd\_win} indicate the fraction of transits with multiple peaks or a non-standard window shape, respectively. Values of 10 or higher indicate a close companion (at $\lesssim$\,1") \citep{2022NatAs...6.1185M,2023A&A...680A..53M}.
Our three targets have these fractions equal to zero or at most 1, indicating no significant contamination. The absence of variability or binarity detection in {\em Gaia} and in the literature, together with the low {\tt ruwe} value, suggests that our targets may be single stars, although the absence of evidence cannot be taken for evidence of absence, especially in the case of binary stars.

\subsection{Spectra preparation}

We downloaded the pipeline-reduced spectra from the ESO archive\footnote{\url{https://archive.eso.org}}. The reduction includes, among others, the bias and flat-field correction, optimal spectrum extraction, and sky subtraction. We removed the telluric absorption bands of H$_2$O and O$_2$ using Molecfit \citep{2015A&A...576A..77S}. We only applied the telluric correction to the UVES 760 setup, as there are no significant atmospheric features below $\simeq$500\,nm. 
We also applied a provisional continuum normalization with low-order splines, using iSpec\footnote{\url{https://www.blancocuaresma.com/s/iSpec}}\citep{2014A&A...569A.111B,2019MNRAS.486.2075B}. 

To align the spectra to laboratory wavelengths, we applied the barycentric corrections from the FITS headers and measured the radial velocity (RV) by cross-correlation with a synthetic template using iSpec.
The template synthetic spectrum was computed 
assuming an effective temperature T$_{\rm{eff}}$\,=\,6000\,K, a surface gravity log$g$\,=\,4.0\,dex, a global metallicity [M/H]\,=\,--1.5\,dex, an alpha enhancement [$\rm{\alpha}$/Fe]\,=\,0.4\,dex, and a microturbulence $\xi$\,=\,1\,km\,s$^{-1}$. 
The atmospheric model was computed with ATLAS\,12 \citep{1970SAOSR.309.....K,1996ASPC..108..160K}. 
\cite{2014A&A...568A..29C} used FLAMES/UVES (R$\simeq$ 47000, wavelength
range $\sim675-1000$\,nm) spectra for our three targets and provided their RV and sulphur abundances.
Our RV (Table~\ref{tab:obs}) differ from those of \citet{2014A&A...568A..29C} by 0.51 -- 0.74\,km\,s$^{-1}$, compatible within the uncertainties.
We display portions of the resulting spectra, ready for analysis, in Fig.~\ref{fig:spec} to illustrate their quality.

\begin{table}[t]
    \centering
    \caption{Measured line broadening and Li feature strength.}
    \begin{tabular}{lcccc}
    \hline \hline
         Star ID & FWHM & EW$_{\rm{Li}}$ & v\,sin$i$ \\
         &[nm]&[nm] & [km\,s$^{-1}$] \\
         \hline
        37934 & 0.013 & 12.1 & $<$6--7 \\
        44606 & 0.008 & 6.5  & $<$6--7 \\
        58195 & 0.009 & 6.3  & $<$6--7 \\
         \hline \hline
    \end{tabular}
    \label{tab:rot}
\end{table}

\subsection{Stellar rotation}

A relation between Li abundance and stellar rotation has been found in various stellar samples, including young clusters and pre-main sequence stars, main sequence and giant stars in open clusters \citep{2018A&A...613A..63B,2020A&A...635L..13A,2023A&A...674A.157T,2024A&A...687A.234C}, and metal-poor halo stars \citep{1996ApJ...456L.115D}. 
The underlying cause for such relations is not completely established, but possible explanations were explored in the literature for different populations, such as stars near the Li-dip, young clusters, or field red giants, including the interplay between rotational mixing and convection, and the tidal interaction with or angular momentum transfer from a companion \citep{2017RSOS....470192S,2017MNRAS.464.1456J,2019ApJ...880..125C}.

Even with our limited sample and spectral quality, the UVES resolution allows us to measure line broadening, and thus projected stellar rotation, down to about 6--7\,km\,s$^{-1}$. 
The instrumental resolution for the employed setup is R\,=\,42310. 
We used IRAF\footnote{IRAF is distributed by the National Optical Astronomy Observatory, which is operated by the Association of Universities for Research in Astronomy (AURA) under a cooperative agreement with the National Science Foundation.} to estimate the Full-Width at Half Maximum (FWHM) and the Li equivalent width (EW) in the region 570--680\,nm, and we found the values reported in Table~\ref{tab:rot}, which are broadly compatible with the spectral resolution and sampling of UVES.

In dwarfs, convection is negligible and the additional broadening is essentially rotational, so we take the line broadening velocity as a rough proxy for v sin$i$.
We used the \texttt{SALVADOR} (Alvarez Garay et al. in prep.) spectral synthesis code, which performs a $\chi^2$ minimization between the observed lines and a grid of synthetic spectra computed on the fly with \texttt{SYNTHE} \citep{sbordone_04,kurucz_05}. We generated the synthetic spectra using the appropriate atmospheric parameters for each star (see Sect.~\ref{sec:res}) over a range of rotational velocities. We selected about 50 strong, isolated, and unsaturated lines spanning the entire observed spectral range above 560\,nm. For each individual line, we fixed the elemental abundance, convolved the synthetic spectrum to match the instrumental resolution, and treated rotation as a free parameter\footnote{Within {\texttt{SALVADOR}, the rotational broadening is applied to the synthetic spectrum already at the instrumental resolution, using the PyAstronomy implementation, which is based on the physical principles detailed in \citet{2008oasp.book.....G}.}}, to find the best-fit solution at the minimum $\chi^2$. As illustrated in Fig.~\ref{fig:rot}, the projected rotational velocities of these stars are well below 6--7\,km\,s$^{-1}$ (i.e., the resolution of the UVES 760\,nm setup) and essentially compatible with no detected rotational broadening.

\subsection{Stellar activity}
\label{sec:actin}

Li-rich giants have been measured to be more chromospherically active than Li-normal stars. \citet{2022ApJ...940...12S} have evidenced this by detecting strong \ion{He}{i}\,1083\,nm absorption in more than half of their Li-rich sample, compared with less than a tenth in the Li-poor red clump sample. It is unknown whether the same holds for the metal-poor main sequence. It has furthermore been proposed that spallation in flares can produce lithium in stars \citep{2000ApJ...534L.207R}, although this is considered unlikely to explain the whole enrichment on its own \citep[Sect.~6.2 in][and references therein]{2017A&A...604A..44M}. In this context, it is worth searching our spectra for chromospheric indicators.

\begin{table}[t]
    \centering
    \caption{Activity indicators, with the characteristics of the bands used for the measurement (see Sect.~\ref{sec:actin}), i.e., the line center, the core FWHM, and the continuum passbands.}
    \renewcommand{\arraystretch}{1.1}
    \resizebox{\linewidth}{!}{
    \begin{tabular}{lcccc}
    \hline \hline
         Line & $\lambda_{\rm{C}}$ & Core & Cont. & Cont. \\
         & [nm] & [nm] & [nm] & [nm]\\
         \hline
        Ca H \& K         & 393.37 & 0.109 & 390.11 $\pm$ 1.00 & 400.11 $\pm$ 1.00 \\
                          & 396.85 & 0.109 & 390.11 $\pm$ 1.00 & 400.11 $\pm$ 1.00 \\
        \ion{He}{i} D$_3$ & 587.56 & 0.040 & 586.90 $\pm$ 0.25 & 588.10 $\pm$ 0.25 \\
        H$_{\rm{\alpha}}$ & 656.28 & 0.025 & 653.78 $\pm$ 0.25 & 658.78 $\pm$ 0.25 \\
        CaT               & 849.80 & 0.100 & 850.825 $\pm$ 0.075 & 858.025 $\pm$ 0.075 \\
                          & 854.21 & 0.100 & 857.800 $\pm$ 0.075 & 861.775 $\pm$ 0.075 \\
                          & 866.22 & 0.100 & 862.035 $\pm$ 0.065 & \\
         \hline \hline
    \end{tabular}
    \label{tab:actin}
    }
    \tablefoot{Ca~\textsc{ii}~H\,\&\,K use triangular bandpasses in the cores \citep{1978PASP...90..267V,1978ApJ...226..379W}; all other cores bandpasses are rectangular.}
\end{table}

We inspected, on the normalized spectra, the features listed in Table \ref{tab:actin}, using the classical Mount Wilson bands \citep{1978ApJ...226..379W,1978PASP...90..267V}. For the H$_{\rm{\alpha}}$ line, we used the bands by \citet{2023Ap&SS.368...63H}. For the calcium triplet (CaT), we measured the line centroids on our spectra with widths of 0.1 nm \citep{2017A&A...605A.113M} and five reference bands \citep[see Table 2 in][]{2016A&A...595A..11L}. The \ion{Na}{i} D doublet was inspected but not used as an index, as its core depth follows the photospheric sodium (Na) abundance, which spans $0.27$ dex among these three stars, most likely because of the Na-O anticorrelation (Sects.~\ref{sec:mp} and \ref{sec:disc}). The narrow positive features observed in two of the H$\alpha$ cores (third panel in Fig \ref{fig:spec}) are most likely cosmic rays. They are narrower (FWHM\,$\simeq$\,0.006-0.009\,nm) than both the instrumental profile ($\simeq$0.016\,nm, at this resolution) and the $\simeq$\,0.047\,nm thermal Doppler width that hydrogen has at $\sim10^{4}$ K, needed to produce the H$\alpha$ line. These spurious features were masked before any index was computed.

We compare our stars because their stellar parameters are similar, but it is unlikely that all three exhibit the same activity levels at the same time, and none of these diagnostics show signs of activity. The \ion{Ca}{ii} K core has an S/N of $\simeq$1--2 per pixel, too low for a reliable use of an $S$ index, but this should not prevent us from seeing a signature --if any-- of a pronounced core reversal, which is common in active stars \citep[e.g. see fig B.1 in][]{2024A&A...684A.106M}. No evident absorption or emission is visible in \ion{He}{i}\,D$_3$, which is commonly used as a flare-sensitive diagnostic \citep{2019PhDT.......127L}, all measures are consistent with noise. The \ion{Ca}{ii} triplet profiles agree reasonably well among the three stars within the uncertainties. They show no evident relative core filling-in, as would be expected for an active star, nor any apparent core emission, which would be indicative of high activity levels \citep{2016A&A...595A..11L}.

\begin{table}[t]
    \centering
    \caption{Adopted parameters for the abundance analysis.}
    \begin{tabular}{lccccc}
    \hline \hline
         Star ID & T$_{\rm{eff}}$ & log$g$ & $\xi$ & [M/H] & [$\alpha$/Fe] \\
         &[K]&[dex]&[km\,s$^{-1}$]&[dex]& [dex] \\
         \hline
        37934 & 5979 & 4.11 & 1.3 & --1.36 & 0.4 \\
        44606 & 5966 & 4.06 & 1.3 & --1.36 & 0.4 \\
        58195 & 5972 & 4.08 & 1.3 & --1.36 & 0.4 \\
         \hline \hline
    \end{tabular}
    \label{tab:pars}
\end{table}

We used H$\alpha$ for an external comparison, after degrading our spectra to the resolution of the comparison survey \citep{2023Ap&SS.368...63H} the three stars lie $5-12$\,\% above the median of a subsample of $95$ main-sequence stars of \citet{2023Ap&SS.368...63H} with [Fe/H]\,$<$\,--1.0\,dex and similar T$_{\rm eff}$ and $\log g$, while the most H$_{\rm{\alpha}}$-active members of that same sample lie $74$\,\% above it. On their bolometric scale (i.e. their Eq. 5, $R_{\rm{H}\alpha}=\Delta\lambda\,\chi\,I_{\rm{H}\alpha}$ with $\chi=f_{\rm cont}/\sigma T_{\rm eff}^{4}\approx1.11\times10^{-4}$\,\AA$^{-1}$ at our parameters) this gives $\log R_{\rm{H}\alpha}\approx-5.07$, for our three stars against $\sim-5.11$ for the median of the matched sample and $-4.87$ for its most active members. These measures can be compared directly on their Figs. 4(a) and 9(a).

\begin{figure*}[t!]
    \centering
    \includegraphics[width=\textwidth]{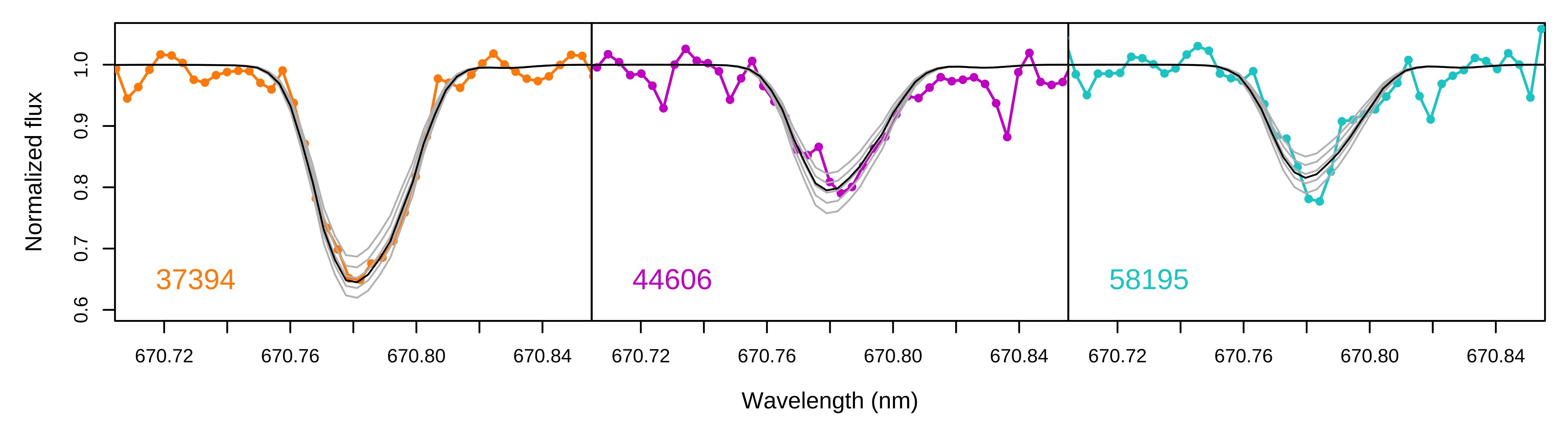}
    \caption{Illustration of the fits of the Li resonant doublet at 670.8\,nm. Each panel displays one of the three targets, as labeled. The best fits are shown as black lines, while a few models with bracketing abundances in steps of 0.05\,dex are plotted in grey.}
    \label{fig:li}
\end{figure*}

We note that residual differences between instruments and pipelines prevent a more accurate comparison between the samples, so we do not attempt to place our estimates on an absolute activity scale. However, this is in any case what is expected at the age of M\,4 for stars in this regime, as for these old stars magnetic braking by stellar wind had enough time to spin them down \citep{1972ApJ...171..565S}, and to deactivate the dynamo that (likely) sustains an active chromosphere \citep{1984ApJ...279..763N}. Moreover, metal-poor stars are expected to be less active at fixed age, because their shallower convective envelopes give them larger Rossby numbers \citep{2020ApJ...889..108A}. Therefore, we conclude that these stars show no evidence of enhanced chromospheric activity.

\section{Abundance analysis}
\label{sec:res}

We carried out the abundance analysis using the code MyGIsFOS \citep{2014A&A...564A.109S}, which derives stellar abundances through spectral synthesis by comparing observed spectra with a precomputed grid of synthetic spectra. 
The grid was computed with SYNTHE \citep{kurucz_05} and ATLAS12 \citep{1970SAOSR.309.....K, 1996ASPC..108..160K} models using the lines selections and solar abundances detailed in \cite{2021A&A...651A..20C, 2024A&A...683A..72C}. 
The grids covers effective temperatures from 5000 K to 6800 K at steps of 200 K, surface gravities from 3.0 to 5.0 at steps of 0.5 dex, metallicities from $-4.00$ to $-0.5$ at steps of 0.50 dex, microturbulent velocities of 0.0, 1.0 and 2.0 $kms^{-1}$, and [$\alpha$/Fe] = $-0.4, 0.0$, and $0.4$. 
To derive the metallicity, we used the T$_{\rm{eff}}$ and log$g$ computed by \cite{2012A&A...539A.157M} for star 37934 and those computed by \cite{2014A&A...568A..29C} for stars 44606 and 58195 as first guesses of stellar parameters (see Table~\ref{tab:pars}). 
For $\xi$, \citep{2016A&A...594A..79S} used 1.3\,km\/s$^{-1}$ and 1.7\,km\/s$^{-1}$ for star 37934. We adopted 1.3\,km\/s$^{-1}$ for all targets, which yields a slope compatible with zero in the A(Fe) versus reduced equivalent width plane when using the \ion{Fe}{i} lines measured in our spectra. Keeping the stellar parameters fixed, we run MyGIsFOS with the derived metallicity to estimate the final chemical abundances. 
We used the Gaia-ESO survey linelist as a starting reference \citep{2021A&A...645A.106H} to identify line masks for MyGIsFOS and also for the subsequent abundance computation. In Table \ref{tab:linelist} (available at the CDS), we provide the linelist adopted. 
Since Li does not scale with metallicity, MyGIsFOS cannot fit the Li line.
For this reason, for Li and those species with only a few, small, or contaminated lines, we used the Python Wrapper for Turbospectrum and Fitprofile (WTF) that generates synthetic spectra with Turbospectrum \citep{1998A&A...330.1109A, 2012ascl.soft05004P} and fits them with Fitprofile \citep{2016A&A...588A..66T}.
We present and discuss the final abundances in the next subsections and report them in Table~\ref{tab:abu}. When we measured more than one line of a given species, we used the mean and the standard deviation ($\sigma$) as the final estimate. To test the sensitivity of the abundance to uncertainties in the atmospheric parameters, we repeated the analysis by altering the parameters by the typical uncertainties (Table~\ref{tab:param_err_abu}). In the following we use abundances in bracket notation\footnote{[Fe/H]\,=\,A$_*$(Fe)\,/\,A$_*$(H)\,--\,A$_{\rm{\odot}}$(Fe)\,/\,A$_{\rm{\odot}}$(H).}, when needed, which were rescaled adopting the \citet{2009LanB...4B..712L} solar reference composition. 

\begin{table}[t!]
    \caption{Abundance ratios of the measured elements, A(El).}
    \centering
    \begin{tabular}{lcccccc}
        \hline \hline
         &star 37934 & star 44606 & star 58195 \\
         &[dex] & [dex] & [dex] \\
         \hline  
\ion{Li}{I}  & 3.01 $\pm$ 0.05 (1) & 2.58 $\pm$ 0.10 (1) & 2.51 $\pm$ 0.10 (1) \\
\ion{C}{I}   & 7.06 $\pm$ 0.14 (3) & 7.37 $\pm$ 0.05 (4) & 7.42 $\pm$ 0.19 (3) \\
\ion{N}{I}   & 7.57 $\pm$ 0.00 (1) & 6.90 $\pm$ 0.00 (1) & 7.26 $\pm$ 0.00 (1) \\
\ion{O}{I}   & 7.87 $\pm$ 0.11 (3) & 8.25 $\pm$ 0.09 (3) & 7.94 $\pm$ 0.12 (2) \\
\ion{Na}{I}  & 5.52 $\pm$ 0.06 (3) & 5.26 $\pm$ 0.15 (4) & 5.34 $\pm$ 0.06 (4) \\
\ion{Mg}{I}  & 6.64 $\pm$ 0.07 (6) & 6.74 $\pm$ 0.08 (7) & 6.74 $\pm$ 0.12 (5) \\
\ion{Al}{I}  & 5.51 $\pm$ 0.06 (2) & 5.57 $\pm$ 0.02 (2) & 5.62 $\pm$ 0.06 (2) \\
\ion{Si}{I}  & 6.69 $\pm$ 0.16 (14) & 6.78 $\pm$ 0.17 (13) & 6.70 $\pm$ 0.20 (12) \\
\ion{S}{I}   & 6.44 $\pm$ 0.06 (2) & 6.49 $\pm$ 0.00 (1) & 6.25 $\pm$ 0.00 (1) \\
\ion{K}{I}   & 4.58 $\pm$ 0.00 (1) & 4.84 $\pm$ 0.00 (1) & -- \\
\ion{Ca}{I}  & 5.31 $\pm$ 0.11 (17) & 5.45 $\pm$ 0.16 (16) & 5.33 $\pm$ 0.14 (13) \\
\ion{Ca}{II} & 5.64 $\pm$ 0.20 (3) & 5.96 $\pm$ 0.12 (3) & 5.87 $\pm$ 0.05 (2) \\
\ion{Sc}{II} & 2.12 $\pm$ 0.12 (5) & 2.19 $\pm$ 0.15 (4) & 2.23 $\pm$ 0.01 (2) \\
\ion{Ti}{I}  & 4.21 $\pm$ 0.10 (4) & 4.10 $\pm$ 0.10 (5) & 4.01 $\pm$ 0.01 (2) \\
\ion{Ti}{II} & 4.12 $\pm$ 0.18 (11) & 4.14 $\pm$ 0.18 (12) & 4.26 $\pm$ 0.16 (8) \\
\ion{V}{i}   & 2.68 $\pm$ 0.00 (1) & 3.06 $\pm$ 0.00 (1) & -- \\
\ion{V}{II}  & 3.01 $\pm$ 0.03 (2) & 2.99 $\pm$ 0.09 (2) & -- \\
\ion{Cr}{I}  & 4.34 $\pm$ 0.03 (3) & 4.53 $\pm$ 0.00 (1) & 4.07 $\pm$ 0.00 (1) \\
\ion{Cr}{II} & 4.15 $\pm$ 0.00 (1) & 4.39 $\pm$ 0.00 (1) & 4.58 $\pm$ 0.00 (1) \\
\ion{Mn}{I}  & 3.99 $\pm$ 0.07 (2) & 4.28 $\pm$ 0.00 (1) & 4.27 $\pm$ 0.00 (1) \\
\ion{Fe}{I}  & 6.06 $\pm$ 0.18 (93) & 6.15 $\pm$ 0.19 (83) & 6.12 $\pm$ 0.24 (74) \\
\ion{Fe}{II} & 6.07 $\pm$ 0.25 (9) & 6.08 $\pm$ 0.28 (6) & 6.07 $\pm$ 0.40 (7) \\
\ion{Co}{I}  & 3.66 $\pm$ 0.00 (1) & 3.39 $\pm$ 0.00 (1) & 3.54 $\pm$ 0.00 (1) \\
\ion{Ni}{I}  & 5.21 $\pm$ 0.08 (3) & 4.93 $\pm$ 0.01 (2) & 4.91 $\pm$ 0.07 (3) \\
\ion{Sr}{II} & 2.06 $\pm$ 0.10 (2) & 2.10 $\pm$ 0.08 (2) & 2.10 $\pm$ 0.19 (2) \\
\ion{Y}{II}  & 1.26 $\pm$ 0.16 (2) & 0.89 $\pm$ 0.17 (2) & 1.69 $\pm$ 0.00 (1) \\
\ion{Zr}{II} & 1.30 $\pm$ 0.03 (2) & 1.81 $\pm$ 0.00 (1) & 1.71 $\pm$ 0.18 (2) \\
\ion{Ba}{II} & 1.47 $\pm$ 0.08 (2) & 1.61 $\pm$ 0.14 (2) & 1.63 $\pm$ 0.19 (2) \\
\ion{Nd}{II} & 0.18 $\pm$ 0.00 (1) & 0.10 $\pm$ 0.00 (1) & 0.80 $\pm$ 0.00 (1) \\
         \hline \hline
         \multicolumn{4}{l}{\tablefoot{For each star and species, we report the abundance plus or minus the standard deviation ($\sigma$, which is 0.00 if only one line is measured), and, in parentheses, the number of lines analyzed. For A(Li), the uncertainty is based on the model grid steps that contain 1\,$\sigma$ of the fit residuals.}}
    \end{tabular}
    \label{tab:abu}
\end{table}

\begin{figure}[t]
    \centering
    \includegraphics[width=\columnwidth]{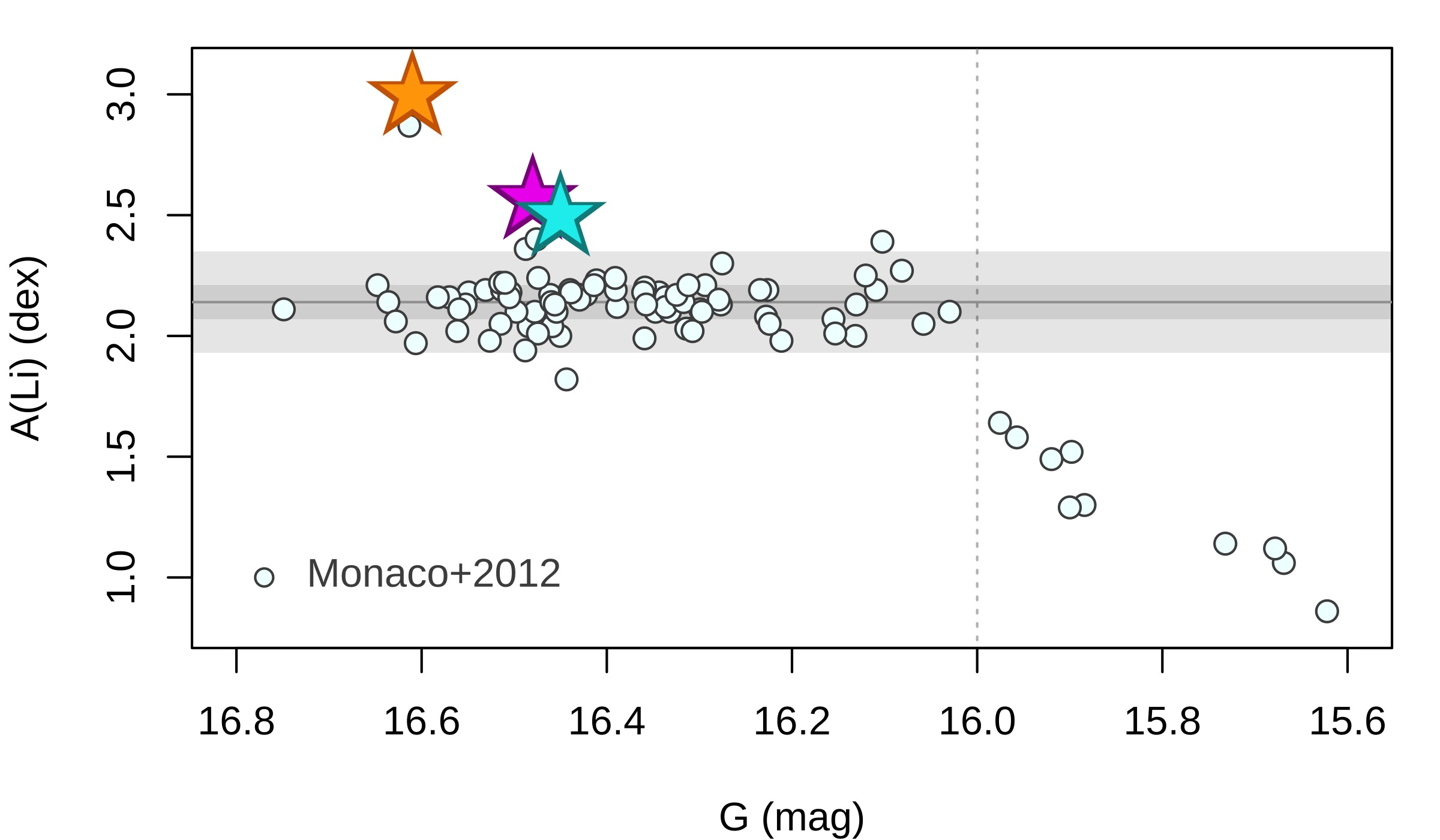}
    \caption{Comparison of the A(Li) of our three targets (symbols colored as in the previous figures) with the \citet{2012A&A...539A.157M} results (empty symbols). The horizontal line shows the median A(Li) from \cite{2012A&A...539A.157M}, with the 1 and 3\,$\sigma$ levels shown as shaded areas. The approximate position of the first dredge-up episode is marked by a vertical dotted line. Our error bars are smaller than the symbols.}
    \label{fig:li2}
\end{figure}

\subsection{Lithium}
\label{sec:lifit}

Table~\ref{tab:abu} lists our computed Li abundances, and Fig.~\ref{fig:li} illustrates the fit. Unlike other element results based on one line, we do provide an A(Li) uncertainty, based on the model grid steps that contain 1\,$\sigma$ of the fit residuals.
The result for star 37934, A(Li)\,=\,3.01\,dex, differs by 0.14\,dex from the abundance reported by \citet{2012A&A...539A.157M}, which is 2.87\,dex, in agreement within uncertainties.
The agreement is particularly remarkable, given that \citet{2012A&A...539A.157M} derived their abundance estimate from the EW of the doublet measured in FLAMES/GIRAFFE spectra using the relations of \citet{2010A&A...522A..26S}, whereas our analysis relies on spectrosynthesis of high-resolution UVES data.
In particular, \cite{2012A&A...539A.157M} reported an EW of 11.40 nm, while we measured EW of 12.05 nm in the UVES spectrum. Applying the NLTE-corrected relationship of \citet{2010A&A...522A..26S} to the EW measured in the UVES spectrum yields A(Li) = 2.94, 
in even closer agreement with the value reported by \cite{2012A&A...539A.157M}. 
We therefore stress that the chemical abundance comparison presented here should be interpreted in light of the possible systematic uncertainties associated with the different datasets and analysis adopted.\\
The other two stars have a smaller A(Li) of about 2.5\,dex, which is higher than the Li plateau (i.e., the Spite plateau \citep{1982A&A...115..357S}) at about 2.2\,dex. 
This conclusion remains unchanged when applying the NLTE-corrected relationship from \cite{2010A&A...522A..26S} to the EW of 6.5 and 6.3\,nm measured in the UVES spectra of stars 44606 and 58195, respectively, yielding A(Li) = 2.47 and 2.43 dex.

The average A(Li) of MS stars measured by \cite{2012A&A...539A.157M} for M\,4 is 2.14\,$\pm$\,0.07\,dex. The abundance of our new Li-rich stars is incompatible with that of normal MS stars at the 6.3 and 5.3\,$\sigma$ level for stars 44606 and 58195, respectively. If we instead consider the large sample by \citet{2009A&A...503..545L} for NGC\,6397, the discrepancy would be at the 4.4 and 3.4\,$\sigma$ level, respectively.
A more homogeneous comparison between the chemical abundances of the Li-rich stars and those at plateau level would help draw stronger conclusions about possible systematic uncertainties. However, the available spectroscopic data do not allow a rigorous, homogeneous analysis.
We thus conclude that these two new stars, 44606 and 58195, are moderately Li-rich, with a Li abundance incompatible with the remaining stars in the \citet{2012A&A...539A.157M} sample, see Fig. \ref{fig:li2}.

\begin{figure*}[t]
    \centering
    \includegraphics[width=\textwidth]{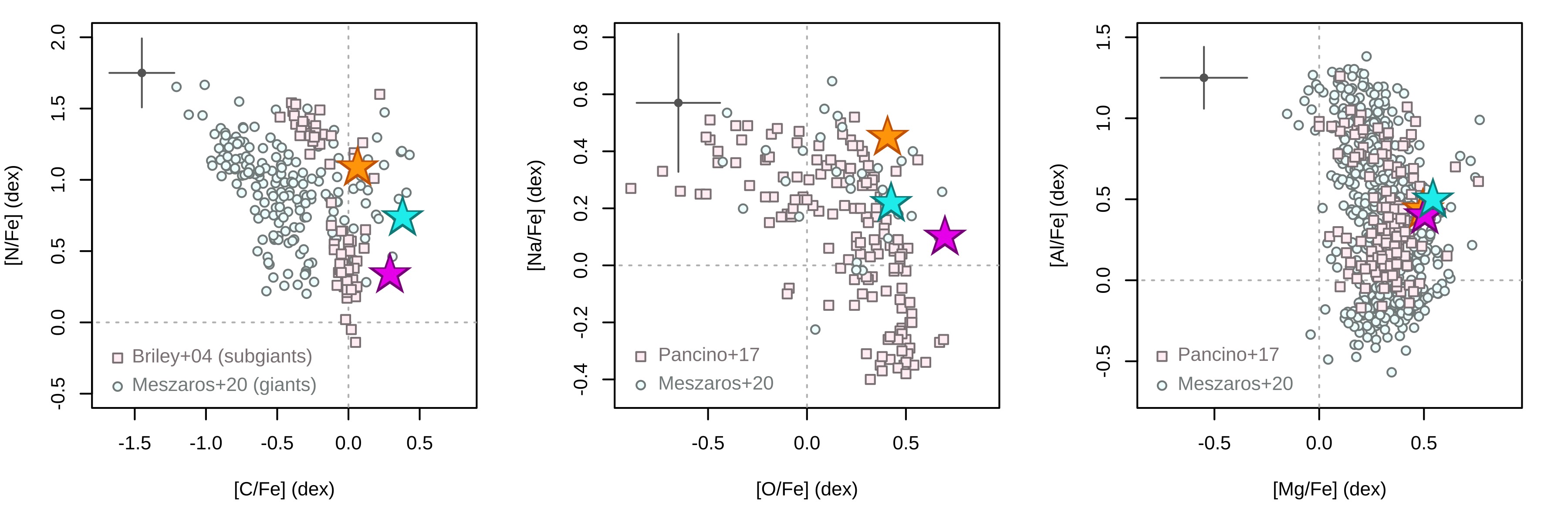}
    \caption{Comparison of the targets abundance with multiple populations in GCs, in the three classical anti-correlation planes: C-N (left), Na-O (center), and Mg-Al (right). Since our O measurements come from the 777\,nm triplet, we corrected our abundances for 3D and NLTE (Sect.~\ref{sec:mp}). Literature samples are plotted in the background in each panel, as labeled: the M\,13 dwarfs by \cite{2004AJ....127.1579B}, the Gaia-ESO Survey giants by \cite{2017A&A...601A.112P}, and all stars with --1.6\,$<$\,[Fe/H]\,$<$--1.1\,dex by  \cite{2020MNRAS.492.1641M}, excluding $\omega$\,Cen. All literature values have been reported to the \cite{2009LanB...4B..712L} solar abundance. Dashed lines mark the solar abundance in all panels. Our targets are plotted as in previous figures. Typical (median) errorbars are plotted in the top-left corner of each panel.}
    \label{fig:mp}
\end{figure*}

\begin{figure}[t]
    \centering
    \includegraphics[width=\columnwidth]{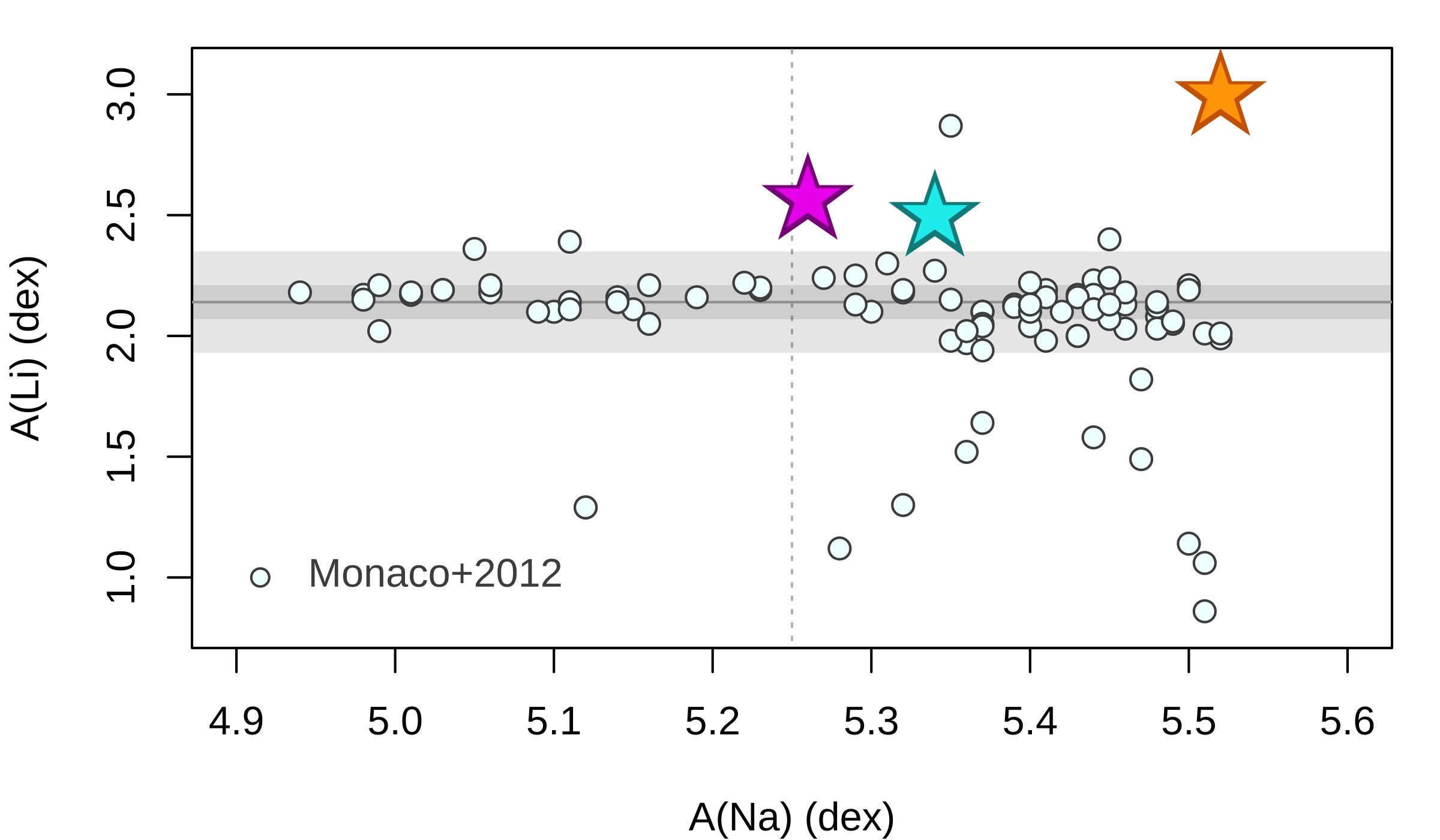}
    \caption{Comparison of the A(Li) of our three targets (symbols colored as in the previous figures) with the \cite{2012A&A...539A.157M} results (empty symbols) as a function of A(Na). The vertical dotted line marks the approximate separation between primordial and enriched populations, obtained from a two-Gaussian fit of the A(Na) distribution by \citet{2012A&A...539A.157M}.}
    \label{fig:lina}
\end{figure}

\subsection{Proton-capture elements and multiple populations}
\label{sec:mp}

We obtain [C/Fe] = 0.06 $\pm$ 0.23 for star 37934, which is 0.28\, dex higher than the value derived by \cite{2016A&A...594A..79S}.
As noted in the following sections, we find a systematic offset of about $\sim$0.2 dex between our chemical abundances and \cite{2016A&A...594A..79S}, which may be due to different data, features and models adopted.
\cite{2016A&A...594A..79S} derived C abundances by fitting the CH G-band between 426 and 431.6 nm in FLAMES-GIRAFFE data using three-dimensional hydrodynamical MARCS model atmospheres \citep{2008A&A...486..951G}, whereas our determination is based on atomic \ion{C}{i} lines (see Tab \ref{tab:linelist}).
The two measurements should therefore not be expected to agree exactly, especially given the different sensitivities of the CH and \ion{C}{i} features to the adopted atmospheric structure and stellar parameters.

We derived nitrogen (N) abundances from the CN molecular band between 388.0 and 388.4 nm, using our C and oxygen (O) abundances to account for molecular equilibrium. 
We compare N and C abundances in the left panel of Fig.~\ref{fig:mp}. The well-known C-N anti-correlation becomes steeper and more C-rich as log$g$ increases, due to extra-mixing as the stars evolve off the main sequence \citep{2004AJ....127.1579B,2012A&A...541A.141L,2020MNRAS.492.1641M, 2025A&A...700A.213V}.
Both the \cite{2016A&A...594A..79S} and our (higher) C abundance determinations are broadly compatible with the expected C abundance for main sequence stars in GCs. 
At the same time, our targets lie in the N-enhanced region of C-N plane.
Their abundance patterns therefore suggest they may belong to the second multiple population, consistent with the sodium (Na) abundance.

We derived the O abundances from the \ion{O}{i} triplet.
We note that the 3D NLTE corrections for the \ion{O}{i} triplet are about --0.13\,dex \citep{2019A&A...630A.104A,2021MNRAS.508.2236B}. Considering that the typical literature values for red  giants reported in Fig.~\ref{fig:mp} are based either on the 630\,nm forbidden line or on infrared features, which do not suffer from significant NLTE effects, we corrected our measurements before plotting them on the Na-O anti-correlation plane. This brings our measurements on the fringes of the classical Na-O anti-correlation to the top-right of the red giant distribution. In that region, a few stars appear in both the \citet{2017A&A...601A.112P} and \cite{2020MNRAS.492.1641M} samples, but no other information about them is given. We tentatively assign our targets to the enriched (or second) population in M\,4 (see Sec.~\ref{sec:mp}), but they might also belong to a different group, with higher O than typical GC stars. 

Applying the NLTE correction by \cite{2007A&A...464.1081A}, we obtain Na abundances of about 5.35\,dex for star 37934 from the Na doublet. 
This is in excellent agreement with the value of A(Na) = 5.35\,dex reported by \cite{2012A&A...539A.157M}, and with the A(Na)= 5.29\,dex derived by \cite{2016A&A...594A..79S}.
As found by \cite{2012A&A...539A.157M} for star 37934, our two new Li-rich stars also show relatively high Na abundance (Fig.~\ref{fig:lina}), supporting our association with the enriched or second multiple population of M\,4 (Fig.~\ref{fig:mp}).

As shown in Fig.~\ref{fig:mp}, the magnesium (Mg) and aluminum (Al) abundances of our targets appear to be compatible with those measured in other GCs stars, thus, we tentatively assign them to a position of mild enrichment, between the primordial (or first) and the enriched (or second) population.

We detected a potassium \ion{K}{i} resonance line in the spectra of stars 37934 and 44606, yielding LTE abundances of A(K) = 4.58 and 4.84\,dex, respectively. These outcomes are consistent with the mean abundance A(K) = 4.60\,dex derived for M\,4 by \cite{2024MNRAS.52712120N}. In contrast, \citet{2020MNRAS.492.1641M} reported a lower mean abundance of A(K) = 4.17\,dex, based on \ion{K}{i} lines in the infrared APOGEE spectra. 
We note that the \ion{K}{i} resonance line is strongly affected by NLTE effects. \cite{2019A&A...627A.177R} report NLTE corrections of order $\simeq -0.59$ dex for stars with stellar parameters similar to those of our targets. Applying this correction would reduce our abundance to A(K) = 3.99 and 4.25\,dex for stars 37934 and 44606, respectively. Once the correction is applied, our K abundances fall in line with those of other Milky Way stars in the Hypatia catalog (Fig.~\ref{fig:afe}) and also with stars in $\omega$\,Cen \citep{2022ApJ...928L..11A}. 
A K-Mg anticorrelation was first discovered in NGC\,2419 \cite{2012ApJ...760...86C} and later in less extreme clusters \citep{2015ApJ...801...68M, 2017A&A...600A.104M}. \cite{2012ApJ...761L..30V} proposed that it arises from proton-capture burning at extremely high temperatures in the progenitors of the polluting material, a process that depletes Mg and enhances K. If M\,4 follows the same pattern, our K abundances support assigning both targets to a slightly enriched (second) population.

\subsection{Alpha elements}
\label{sec:alpha}
For star 37934, we obtain silicon over iron of [Si/Fe] = 0.56 $\pm$ 0.24\,dex, in agreement within errors with the value of [Si/Fe] = 0.61\,dex reported by \cite{2016A&A...594A..79S}.

\cite{2014A&A...568A..29C} derived Sulphur (S) abundances for the stars analysed in this work, using \ion{S}{i} lines from multiplet (Mult.) 8 at 675 nm and Mult. 1 at 920 nm.
They found A(S) = 6.38 $\pm$ 0.08, 6.28 $\pm$ 0.13, and 6.25 $\pm$ 0.07\,dex for stars 37934, 44606, and 58195, respectively.
Using the S I lines of Mult. 1, we obtained A(S) in agreement within uncertainties: A(S) = 6.44 $\pm$ 0.06 for 37934, A(S) = 6.49 $\pm$ 0.00\,dex for 44606, and A(S) = 6.25 $\pm$ 0.00\,dex for 58195.

\begin{figure}[t]
    \centering
    \includegraphics[width=\columnwidth]{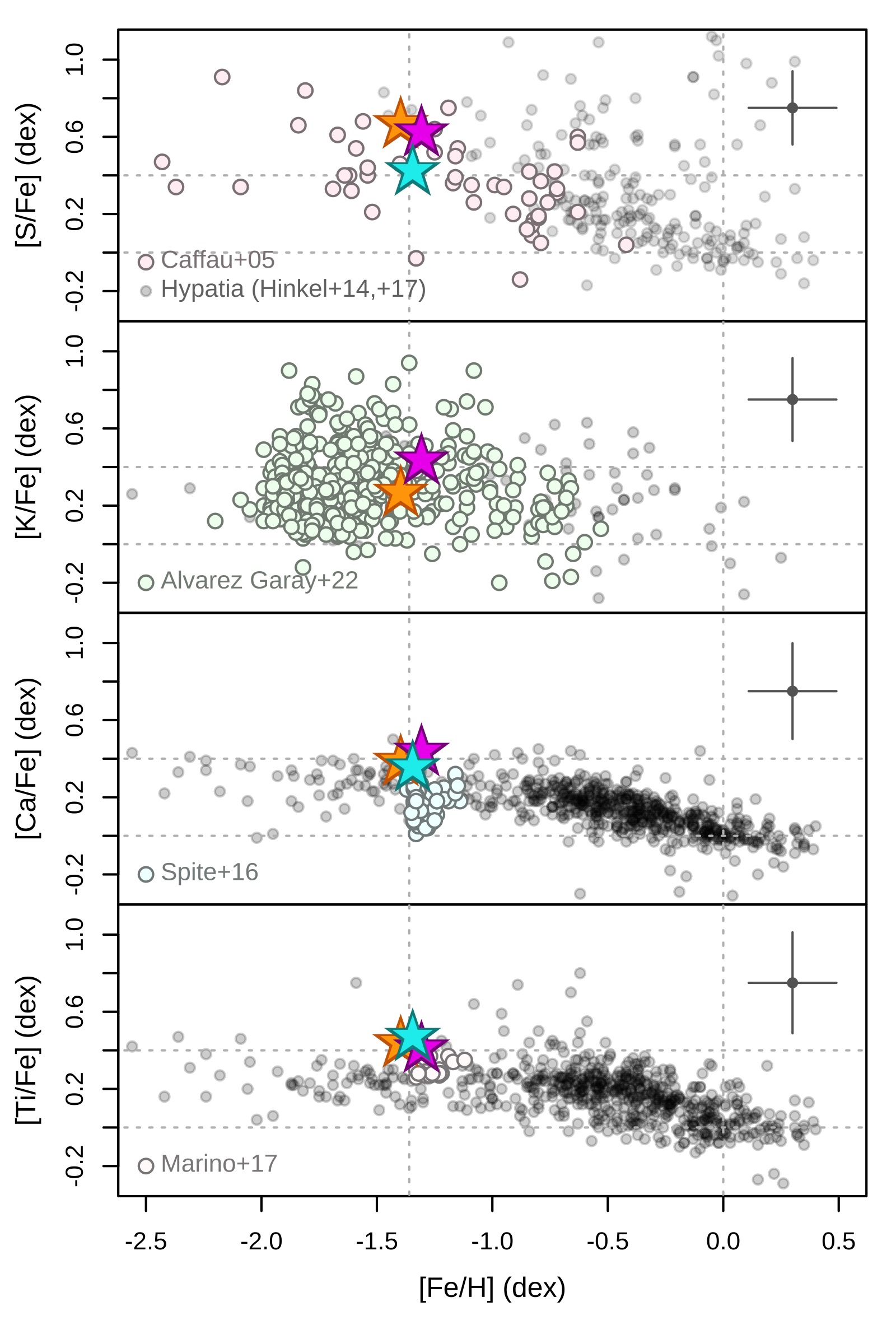}
    \vspace{-0.6cm}
    \caption{Example of $\alpha$-element results. In each panel, horizontal dashed lines show the solar and $\alpha$-enhanced abundance ratios, vertical lines show the solar and M\,4 metallicity. Grey and empty symbols in the background represent literature reference Milky Way samples from the Hypatia catalog \citep{2014AJ....148...54H,2017arXiv171204944H}, from \citet{2005A&A...441..533C}, $\omega$\,Cen measurements by \citet{2022ApJ...928L..11A}, and M\,4 measurements from \citet{2016A&A...594A..79S} and \cite{2017ApJ...843...66M}. Literature values are reported to the \citet{2009LanB...4B..712L} solar reference abundance. Our measurements are reported as in previous figures, corrected for NLTE effects (Sect.~\ref{sec:alpha}). Typical (median) errorbars are reported in the top-right corner of each panel.}
    \label{fig:afe}
\end{figure}

For star 37934, we obtain calcium over iron of about 0.39 $\pm$ 0.21\,dex, which is 0.23 dex higher than the value obtained by \cite{2016A&A...594A..79S}. This difference is comparable to our quoted uncertainty and therefore does not indicate a significant discrepancy, especially given the excellent comparison with the Milky Way distribution (Fig.~\ref{fig:afe}).
Our result agrees with the mean cluster abundance of [Ca/Fe] = 0.40 dex derived by \cite{2011A&A...535A..31V}.
More recent analyses report lower mean values, with [Ca/Fe] = 0.27 - 0.32\,dex for M\,4 \citep{2017A&A...601A.112P, 2020MNRAS.492.1641M}, but the difference is well within uncertainties.

\cite{2020MNRAS.492.1641M} found no significant difference in titanium abundance between the N-rich and N-poor stellar populations, with mean abundances of A(Ti) = 4.15 and 4.10\,dex, respectively. The mean Ti abundances derived for M\,4 by \cite{2011A&A...535A..31V} is A(Ti) = 4.09\,dex. Thus, the Ti abundances measured for the three Li-rich stars are consistent with the typical Ti abundance of M\,4.

\begin{figure}[t]
    \centering
    \includegraphics[width=\columnwidth]{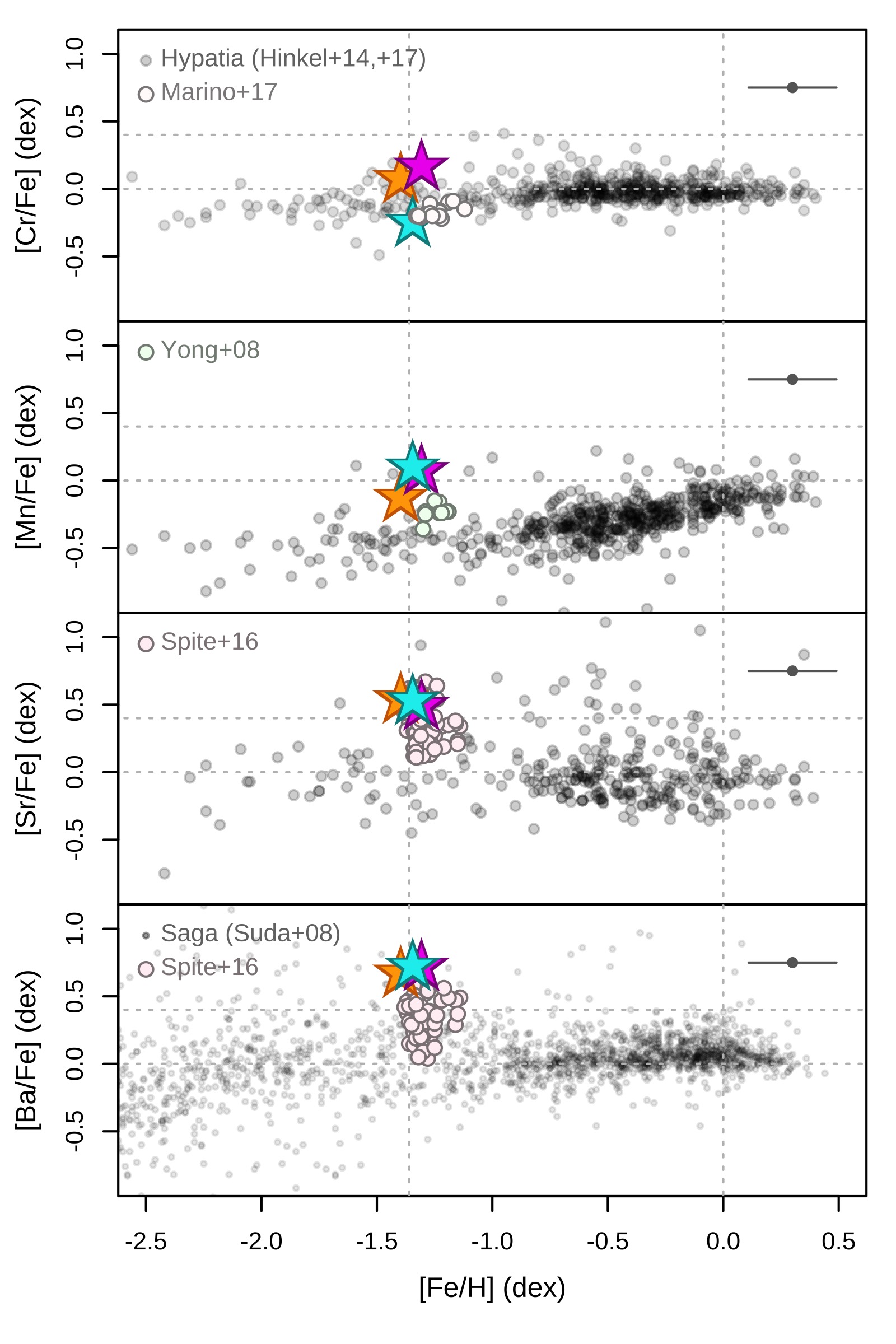}
    \vspace{-0.6cm}
    \caption{As Fig.~\ref{fig:afe} but for the iron-peak and s-process elements results.}
    \label{fig:iron}
\end{figure}

\subsection{Iron-peak elements}
\label{sec:iron}

A total of 95 \ion{Fe}{i} and 9 \ion{Fe}{ii} lines were detected and considered to derive the metallicity in our spectra. We find a good agreement between A(\ion{Fe}{i}) and A(\ion{Fe}{ii}), with no indication of a systematic offset among the three stars.
\cite{2012A&A...539A.157M} found two metallicity values for M\,4, namely [Fe/H]$=-1.30$ dex for hot MS stars (T$_{\rm{eff}}>5800$ K) and [Fe/H]$=-1.17$ dex for cooler, brighter stars (T$_{\rm{eff}}<5600$ K). 
They attributed this difference mainly to the adopted microturbulence velocities and considered the values compatible, within the uncertainties, with previous estimates, such as  [Fe/H]$=-1.10$ dex from \cite{2011MNRAS.412...81M}.
 For star 37934, \cite{2012A&A...539A.157M} derived A(Fe) = 6.16\,dex, a value subsequently adopted by \cite{2014A&A...568A..29C} in their analysis of stars 37934, 44606, and 58915. Our measurements, A(Fe)= 6.06 $\pm$ 0.18, 6.15 $\pm$ 0.19, and 6.12 $\pm$ 0.24\,dex for stars 37934, 44606, and 58915, respectively, agree well with these determinations.
Previous spectroscopic studies consistently find a small dispersion in iron abundance across different evolutionary stages, with mean metallicities for M\,4 ranging from [Fe/H]\,$\simeq-1.1$ \citep{2011ApJ...730L..16M,2020MNRAS.492.1641M} to [Fe/H]\,$\simeq -$1.2\,dex \citep{2008A&A...490..625M, 2009A&A...508..695C, 2016A&A...594A..79S}. Overall, our metallicities are consistent within the uncertainties with the established metallicity of M\,4 \citep{1996AJ....112.1487H,2010arXiv1012.3224H}.

We derive a mean value of scandium over iron, [Sc/Fe]\,$\simeq\,$0.46 dex for M\,4, which is higher than the values reported in the literature around [Sc/Fe]\,$\simeq$\,0.15 dex \citep{2008ApJ...689.1031Y, 2017ApJ...843...66M}. However, \cite{2008ApJ...689.1031Y} and \cite{2017ApJ...843...66M} are based on RGB and asymptotic giant branch (AGB) stars, respectively, which may explain the marginal discrepancy.

We detected vanadium lines only in stars 37934 and 44606, defining a mean [V/Fe]\,$\simeq$\,0.23 dex for M\,4, and confirming the value found by \cite{2008ApJ...689.1031Y}.

Our measurements confirm a mean chromium abundance close to the solar value, as previously established for the behavior of this element in M\,4 \citep{2008ApJ...689.1031Y, 2017ApJ...843...66M}.

Cobalt is also approximately solar relative to iron, with no significant star-to-star variation. Our results are consistent with previous measurements in M\,4, which reported mean [Co/Fe] ratios of $\sim$0.01 dex \citep{2008ApJ...689.1031Y} and $-$0.04 dex \citep{2017ApJ...843...66M}.

The nickel abundances of our sample are consistent with the value A(Ni)=4.99$\pm$0.04\,dex for MS stars at TO point in M\,4 \citep{2024MNRAS.52712120N}.
For star 37934, we find [Ni/Fe] = 0.39 $\pm$ 0.19 dex. \cite{2016A&A...594A..79S} derived [Ni/Fe] = 0.18\,dex from \ion{Ni}{i} line at 676.77\,nm.
Our A(Ni), determined from the \ion{Ni}{i} line at 676.77 nm, differs by only 0.12 dex from \cite{2016A&A...594A..79S}.

All these abundances appear to be roughly solar within uncertainties, as expected. The example of Cr is shown in Fig.~\ref{fig:iron}.

Finally, we also present the abundance of manganese, an element rarely measured in GCs. 
We derive [Mn/Fe]\,=\,$-0.11$ dex for star 39734, in good agreement within the uncertainties with the value of $-0.24$\,dex reported by \cite{2008ApJ...689.1031Y} for M\,4. Interestingly, the other two Li-rich stars show a higher [Mn/Fe] ratio around $\sim$0.1 dex. However, previous studies have shown that NTLE corrections for Mn I can be of order 0.1 dex for stars with stellar parameters similar to those of our targets \citep{2020A&A...643A..49H}.

\subsection{Neutron-capture elements}
While stars 44606 and 58915 show zirconium abundance in agreement with the results obtained by \cite{2008ApJ...689.1031Y}, [Zr/Fe] = 0.48 $\pm$ 0.04\,dex, star 37934 appears to be relatively Zr-poor, with [Zr/Fe] = 0.12 $\pm$ 0.24\,dex.

For star 37934, our result of strontium over iron, [Sr/Fe] = 0.56 $\pm$ 0.21\,dex, differs by 0.29\,dex from \cite{2016A&A...594A..79S}. 

We derived yttrium abundances and found a mean value of [Y/Fe]\,$\simeq 0.42$ dex, in agreement with the enhanced Y abundances previously reported for M\,4 \citep{2008ApJ...689.1031Y, 2013MNRAS.433..366D}.

Our result for star 37934 of barium over iron, [Ba/Fe] = 0.69 $\pm$ 0.19\,dex, is in good agreement with \cite{2007A&A...464.1081A}, [Ba/Fe] = 0.64\,dex.
The other two Li-rich stars show a slightly higher A(Ba), but agree with the literature within error.

A neodymium line, \ion{Nd}{ii}, was detected and used in this work.
\cite{2008ApJ...689.1031Y} found [Nd/Fe] values between 0.43 and 0.64 dex. Our results indicate that star 58195 ([Nd/Fe] = 0.73\,dex) shows the Nd enhancement expected in M\,4, whereas 37934 and especially 44606 are significantly lower. These measurements are based on one line and the discrepancy between our targets suggests that these results must be taken with caution.

\section{Discussion}
\label{sec:disc}

We now discuss the properties of our three targets, presented in Sects.~\ref{sec:data} and \ref{sec:res}, in light of the proposed interpretations of the origin of Li-rich stars, briefly introduced in Sect.~\ref{sec:intro}.

\subsection{Interplay with multiple populations}\label{sec:mp_interplay}

We start by stressing that, as illustrated in Sect.~\ref{sec:mp} (see also Fig.~\ref{fig:mp}), GCs are affected by the MP phenomenon \citep{2015MNRAS.454.4197R,2018ARA&A..56...83B} and the interplay between the two phenomena complicates the interpretation. 
In other words, if a star in a GC is Li-rich and also shows high Na abundance, Na abundance cannot unambiguously identify the Li-enhancement mechanism. 
A second-population star is expected to be Li-poor and Na-rich, as Na can be produced through proton-capture nucleosynthesis at temperatures higher than those required for Li destruction.
This scenario is consistent with the Li-Na anti-correlation observed in some GCs (\citealt{2009A&A...503..545L} in NGC\,6397; \citealt{2012A&A...539A.157M} in M\,4, \citealt{2015MNRAS.449.4038D} in NGC\,1904, NGC\,2808, NGC\,362; \citealt{2018A&A...618A.134M} in $\omega$ Cen), although the strength and presence of this relation may vary among clusters.
Being Li and Na-rich, our targets break this pattern. Therefore, Li content cannot be a simple consequence of population membership and cannot be explained by the standard MP scenario. It points to a Li-enrichment mechanism acting in addition to the polluters that shaped the Na enrichment.
On the other hand, an enhancement of Li and Na can be related to unmixed material from AGB stars, while most of the Na-rich stars are likely formed from material diluted with pristine gas.\\

The C+N+O abundance is higher than solar, so our targets may belong to a separate population, more similar to the anomalous stars found in several clusters: $\omega$\,Cen \citep{2000ApJ...534L..83P,2011ApJ...730L..16M,2012ApJ...746...14M}, NGC\,1851 \citep{2012A&A...541A.141L}, and M\,2 \citep{2012A&A...548A.107L,2013MNRAS.433.1941L}. These anomalous populations normally have higher than solar s-process abundances, which seems to be the case of our targets, but the whole cluster seems to be s-process enriched \citep{2008ApJ...689.1031Y,2016A&A...594A..79S,2017ApJ...843...66M}. Additionally, M\,4 does not seem similar to more complex GCs, which show anomalous populations and s-process enrichment. We conclude that the relatively high s-process abundance may relate to the unknown Li-enhancement mechanism or to the peculiar chemistry of M\,4 as a whole.

Finally, our targets do not seem to belong to the extreme population \citep[as defined by][]{2009A&A...508..695C}, as illustrated in Fig.~\ref{fig:mp}. This is supported by the fact that in GCs, the Li abundance correlates with MPs, in the sense that the extreme population normally has lower Li abundance than the remaining MPs \citep{2014ApJ...791...39D,2015MNRAS.449.4038D}, although a similar behavior was not clearly detected in M\,4 \citep{2010ApJ...716L.166D}.

\subsection{Sub-stellar companion engulfment}

As mentioned in Sect.~\ref{sec:intro}, in the case of GCs the planet (or sub-stellar companion) engulfment hypothesis can be safely discarded, mostly because planets are rare in metal-poor stars \citep{1997MNRAS.285..403G} and even rarer in GCs. To date, only one planet has been found in a GC compared to about 4\,000 planets known in total \citep{2019MNRAS.489.4311C}. The planet host is a pulsar in M\,4, much fainter than our targets, with a white dwarf companion and a third companion, which is most likely a planet \citep{1993Natur.365..817B}. Seven additional planet candidates were reported in M\,4 \citep{2020AJ....159..106W}, but they are most likely false alarms and none of their host stars are similar to our targets. Notably, planet engulfment should also lead to a high beryllium abundance \citep{2025A&A...697A..75M}, produced by spallation in icy exoplanets or exomoons \citep{2021ApJ...907L..35D}. Unfortunately, Be is out of the wavelength range of our spectra and our targets are too faint to measure it in the near future. As discussed extensively in the literature \citep[see e.g.,][]{1999MNRAS.308.1133S}, the frequency of brown dwarfs does not drop at low metallicity, and brown dwarfs would deposit their entire, undepleted Li content into the envelopes of the engulfing stars. However, according to the computations by \citet{2016ApJ...829..127A}, the maximum surface A(Li) abundance that can be obtained is 2.2\,dex, therefore, brown dwarf engulfment is not sufficient to explain our observations.

\begin{figure}[t]
    \centering
    \includegraphics[width=\columnwidth]{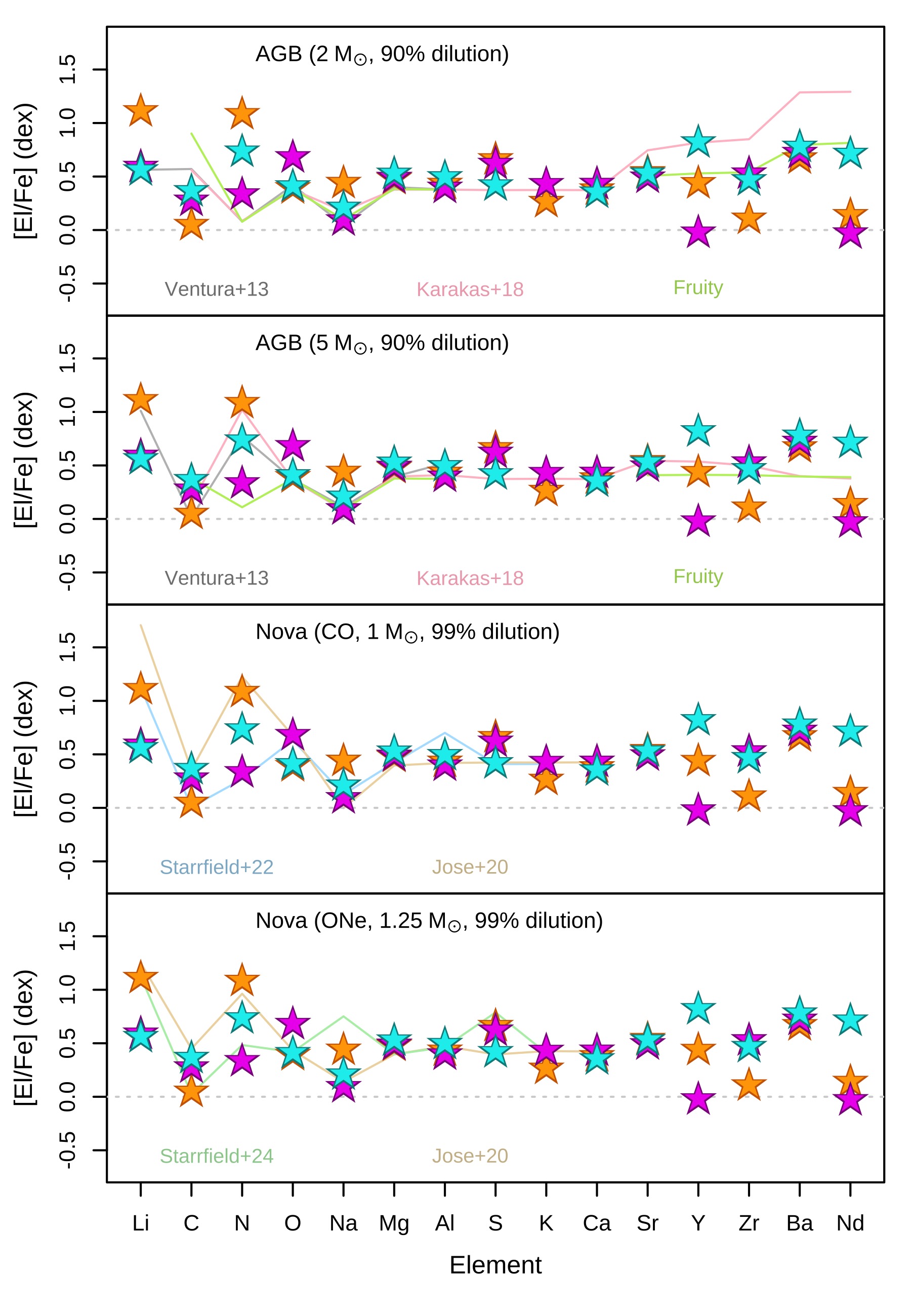}
    \caption{Comparison of our targets abundances (colored symbols) with theoretical predictions from various literature sources. For AGB stars (two upper panels), we compare with the models by \citet{2006NuPhA.777..311S,2007ApJ...667..489C,2009ApJ...696..797C,2013MNRAS.431.3642V,2018MNRAS.477..421K}. For the Nova models, we compare with the CO and ONe models (two lower panels) by \cite{2020ApJ...895...70S,2024ApJ...962..191S} and \cite{2020A&A...634A...5J}. We calibrate the Li yield with the observed value by \citet{2022MNRAS.509.3258M} for RS\,Oph.}
    \label{fig:yields}
\end{figure}

\subsection{Lithium production in red giants}

Lithium can be produced in AGB stars through the Cameron-Fowler mechanism \citep{1971ApJ...164..111C}, which allows for fresh Li production through the 3$\alpha$ reaction and subsequent electron capture on the produced Be, if some extra-mixing mechanism can bring Li up in cooler layers, where it can survive. Depending on the AGB star's mass, production can occur in different phases, from early AGB to thermal pulses \citep{2000A&A...359..563C}. In lower-mass stars with a degenerate core during the RGB, Li can be produced at the RGB bump, where extra-mixing processes connect the $^3$He-rich envelope to hotter internal zones, activating the Cameron-Fowler mechanism and bringing freshly produced Li to the surface. Lithium can also be produced during the helium flash, at the top of the RGB evolution of low-mass stars \citep{2020NatAs...4.1059K}.

Our targets are on the main sequence, therefore, they cannot produce Li through these mechanisms. However, a more massive binary companion, previously evolving through the AGB or RGB-bump phases, could have transferred polluted material onto their surface before evolving into a much fainter object. If the polluter was on the late AGB phase during mass transfer, we would expect an increased C+N+O and/or s-process abundance, depending on its mass \citep{1999ARA&A..37..239B}.  We do observe s-process element abundances higher than those of typical halo stars in our targets. This is especially true for the two new targets, and it is consistent for all the s-process elements measured. However, M\,4 seems to have an overall s-process abundance slightly higher than solar \citep{2008ApJ...689.1031Y,2016A&A...594A..79S,2017ApJ...843...66M}, although our measurements do seem to lie on the upper envelope of the literature measurements. If the s-process enhancement -- with respect to the rest of M\,4 stars -- is considered significant, then AGB models are required, as Novae and other polluters are not expected to produce s-process elements. Both low- and high-mass AGB stars reproduce Na, Mg, and Al reasonably well, if we consider a $\simeq$90\% dilution with pristine gas. However, reproducing the mild Li enhancement in the two new Li-rich stars and reconciling C, N, and O abundances remain difficult, regardless of the chosen AGB mass. Examples of the comparison with 2 and 5\,M$_{\rm{\odot}}$ are shown in Fig.~\ref{fig:yields}.

As described in Sect.~\ref{sec:data}, we searched the literature for evidence of binarity for our targets, but we found none. Therefore, either the Li enhancement is not caused by binary interaction, or we missed evidence for binarity. The reasons may differ, e.g., inadequate monitoring, faint companions, or disruption by the dynamically active environment of M\,4. 

\subsection{Lithium connection with stellar activity and rotation}

Fast rotating stars exhibit higher surface A(Li) compared to slowly rotating ones of similar \teff\, and mass \citep{2002AJ....123.2703D, 2018A&A...613A..63B, 2020A&A...635L..13A, 2024A&A...687A.234C}. Rapid rotators have strong, dynamo-generated magnetic fields that inhibit convective energy transport and inflate the radius. As a consequence, the temperature at the base of the outer convective zone decreases, delaying or suppressing Li destruction. In contrast, slow rotators develop stronger internal differential rotation between the core and envelope, enhancing mixing and Li depletion \citep{2017MNRAS.464.1456J, 2018A&A...613A..63B}.

Rapid surface rotation is frequently induced by tidal interaction in binary systems, where the angular momentum transfer triggers the production of Li through the Cameron-Fowler mechanism \citep{2019ApJ...880..125C}.
Engulfment of a planet or substellar companion is another way to transfer orbital angular momentum to the star's convective envelope. Consequently, the spin-up drives a dynamo that boosts X-ray/magnetic activity while simultaneously depositing external Li-rich material or triggering mixing and Li production \citep{2017A&A...604A..44M}. 
Subsequently, a spun-up star's surface rotation slows due to magnetic braking. Since stars lose their fast rotation much faster, $\sim 10^5$ yr, than they destroy their newly produced Li, $\sim 10^6$ yr, only 10$\%$ of binary-stimulated Li-rich stars are observed with high surface rotation rates \citep{2019ApJ...880..125C}.
This scenario may explain why our stars do not show high rotation, but the assumption of spin-up due to a binary system or tidal interaction is currently unsupported.

Highly active stars display fast rotation, strong magnetic fields, high X-ray flux, asymmetric H$\alpha$ profile emissions (indicative of mass loss), and far-IR excesses from ejected shells \citep{2002AJ....123.2703D}.
Li-rich stars are chromospherically active with a strong \ion{He}{I} 1083 nm absorption line \citep{2022ApJ...940...12S}. 
During energetic flares, low-energy spallation reactions can produce $^6$Li and $^7$Li directly in atmospheric layers.
\cite{2007A&A...469..265T} demonstrated that the cumulative effect of stellar flares over $\sim$10 Gyr of main sequence evolution can produce significant amounts of $^6$Li (and $^7$Li), with $^6$Li/H potentially reaching detectable levels of $\sim10^{-11}$ after several Gyr under favorable conditions. However, this process is insufficient by itself to account for large total surface Li enhancements \citep{2000ApJ...534L.207R}.
As already mentioned, the Li-rich stars presented in this work do not show enhanced chromospheric activity. 
In conclusion, we can exclude rotation and stellar activity as mechanisms of Li production in our targets, unless the stars were active in the past, perhaps due to a binary interaction that spun them up, and they have since spun down by magnetic braking, retaining their enhanced Li abundances.

\subsection{Lithium enhancement in binaries, novae, and mergers}

Besides pollution by a companion that produces Li, other mechanisms can induce Li enhancement. The first is by dynamical processes such as tidal interaction or tidal locking \citep{2019ApJ...880..125C}. Tidal spin-up appears to be the favored mechanism for metal-rich field stars, which are predominantly in the red clump phase. The idea is that the extra rotation provides the extra mixing needed for the Cameron-Fowler mechanism to operate effectively, so the star becomes Li-rich on the red clump. It is unclear whether this explanation applies to our targets, because they are still on the main sequence. 

The second mechanism is stellar mergers \citep{2013MNRAS.430.2113Z,2020ApJ...889...33Z}, where Li can be produced because the merging drives convection and triggers the Cameron–Fowler mechanism if sufficient hydrogen is brought into the core. An interesting result by \citet{2023A&A...672A.196K} showed a moderate Li enhancement in three red novae. This hypothesis, however, works only for the merger of a red giant and a smaller object, such as a white dwarf, resulting in a red giant or at least a subgiant, and produces A(Li)\,$>$\,3\,dex, so it is not applicable to our case. 

The third possibility is novae \citep{2025A&A...699A.171M}, specifically invoked to explain unevolved stars like our targets, which cannot result from the Cameron-Fowler mechanism described above. 
In this framework, thermonuclear runaways occurring on a massive ONe white dwarf ($\simeq 1.2-1.35\,M_{\rm{\odot}}$) synthesize extreme amounts of Li along with high [N/Fe] and [Na/Fe] ratios, while leaving not enhanced C, Al, and s-process elements abundances. A tiny accretion of this nova ejecta onto a companion star, such as an outer tertiary in a hierarchical triple system, is sufficient to raise the target star's surface abundance to A(Li) $\sim\,3$ dex without requiring any internal nuclear processing or post main sequence mixing. This configuration can leave either a single unevolved Li-rich star, if the inner binary subsequently merged or unbound, or a star with a massive compact companion, providing a comprehensive framework that fits the apparent single nature of our targets. We compare our abundances with Novae models in Fig.~\ref{fig:yields}. Following \cite{2025A&A...699A.171M}, we use the yields by \citet{2020ApJ...895...70S, 2024ApJ...962..191S} and \cite{2020A&A...634A...5J}. We use a 25\%--75\% mixing, and we further dilute the ejecta with pristine gas, observing that just 1\% of novae ejecta are sufficient to reproduce the observed chemistry. Since the Li predictions are at least one order of magnitude lower than what is observed in Novae, we use the value observed by \citet{2022MNRAS.509.3258M} for RS\,Oph. The models compare as well as the AGB ones, if not better, especially for C and O. The s-process elements are not expected from novae, but we remind that all stars in M\,4 are s-process enriched and a possible explanation is a peculiar pre-enrichment of the cloud from which M\,4 formed \citep{2013MNRAS.433..366D, 2014ApJ...795...34S}.

As an additional fourth possibility, \cite{2024MNRAS.527.7005K} invoke the merger of hard, high-mass-ratio MS binaries to form second-generation stars in globular clusters, citing \cite{2023A&A...672A.196K} as evidence that such mergers activate the Cameron-Fowler mechanism rather than destroying surface Li, thereby allowing second-generation merger products to retain normal or high Li without pristine-gas dilution. 
Mergers plausibly dominate in dense cluster environments, while novae pollution better explains the unevolved, field Li-rich stars, making these two frameworks complementary in explaining the origin of these objects. This suggests that mergers may plausibly explain the Li enhancement observed in our targets.

\section{Conclusions}
\label{sec:concl}

We have detected two new Li-rich dwarf stars in the GC M\,4, with A(Li) = 2.58 $\pm$ 0.10 and 2.51 $\pm$ 0.10 dex. This brings the number of known Li-rich stars in GCs to 18. Together with 37934, for which we obtain A(Li) = 3.01 $\pm$ 0.05 dex, our targets form a sample of three Li-rich turnoff stars in a single GC. The two new stars lie about 0.3\,dex above the Spite plateau, which is from 3 to 6\,$\sigma$ out of the distribution (Sect.~\ref{sec:lifit}), and are thus Li-rich with respect to the other M\,4 dwarfs studied by \cite{2012A&A...539A.157M}.

We determined their abundance pattern thanks to high-resolution UVES spectra covering a wide spectral range, from $\simeq$\,328 to 946\,nm. The $\alpha$-elements, iron-peak, and most neutron-capture abundances of the three stars are consistent with typical M\,4 values, with only a few marginal differences from the literature.
The C, N, O, Na, Mg, and Al abundances of the Li-rich stars seem to place them in regions of the corresponding anti-correlation planes of enriched (or second) population.

The Li-rich stars 37934, 44606, and 58195 show no measurable rotational broadening (v\,sin$i$ below 6--7\,km\,s$^{-1}$) or chromospheric activity. We can therefore exclude rotation- and flare-driven spallation as the origin of the Li excess, unless the stars were spun up long ago by a companion and are now quiet but retain their enhanced Li abundances. Our target stars are on the main sequence turn off, so they cannot have produced the Li themselves through the Cameron–Fowler mechanism. Planet engulfment is very unlikely in a GC, and engulfing a brown dwarf cannot reach the observed A(Li). 

By elimination, the most plausible explanations involve some form of interaction with a companion. This could be mass transfer from a low-mass AGB star, 
consistent with the mildly enhanced s-process abundances, pollution by nova ejecta, or a merger. However, we find no trace of a companion in the Gaia astrometry or in the literature. This may be due to a lack of adequate monitoring, to faint or compact companions, to disruption of the binary in the dense cluster environment, or to a merger. To reliably discriminate among these scenarios, additional abundances, especially of Be and N, would be useful, as would an analysis of the \ion{He}{i} line at 1083\,nm. Most importantly, it would be necessary to thoroughly search for a binary companion, if present, using radial-velocity monitoring, UV excess, and similar methods.

\begin{acknowledgements}
This paper is based on data collected with the Very Large Telescope (VLT) at the European Southern Observatory (ESO) on Paranal, Chile, under ESO Program ID 089.D-0062(B).
We acknowledge the support of the {\em ESO Visitorship Programme} of ESO Vitacura, Chile, where a large part of the data analysis and interpretation took place. 
Co-funded by the European Union (ERC-2022-AdG, {\em "StarDance: the non-canonical evolution of stars in clusters"}, Grant Agreement 101093572, PI: E. Pancino). Views and opinions expressed are, however, those of the author(s) only and do not necessarily reflect those of the European Union or the European Research Council. Neither the European Union nor the granting authority can be held responsible for them. 
L.M. gratefully acknowledges support from ANID FONDECYT Regular Project n.1251809. 
S.V. gratefully acknowledges the support provided by the ANID BASAL project FB210003. 
\end{acknowledgements}

\bibliographystyle{aa}
\bibliography{biblio}

\begin{appendix}
\section{Additional tables}  

Here we present additional tables related to the abundance analysis. Table~\ref{tab:linelist} shows the linelist used in the abundance analysis, while Table~\ref{tab:param_err_abu} shows how abundances vary with typical changes in the input stellar parameters.


\begin{table}[h!]
    \caption{Linelist adopted in this work.}
    \centering
    \begin{tabular}{lcccccc}
        \hline \hline
        Element&Wavelength&log\,$gf$& $ \chi_{exc}$\\
        &[nm]&&&\\
        \hline \hline
        \ion{Li}{I} & 670.78 & 0.000 & -0.309 \\
        \ion{C}{I} & 906.14 & -0.346 & 7.483 \\
        \ion{C}{I} & 908.85 &-0.429 & 7.483& \\
        \ion{C}{I} & 909.48 & 0.151& 7.488 & \\
        \ion{C}{I} & 911.18 & -0.296&7.488 & \\
        ...&&&&\\
        \hline \hline
    \end{tabular}
    \label{tab:linelist}
\end{table}

\begin{table}
    \caption{Sensitivity of abundances to atmospheric parameters.}
    \centering
    \begin{tabular}{lcccccc}
        \hline \hline
        &$\Delta$T$_{\rm{eff}}$ & $\Delta$log$g$ & $\Delta\xi$ & $\Delta$[M/H] \\
        &$\pm$100 K & $\pm$0.1 & $\pm$0.1 & $\pm$0.1 \\
         &[dex] & [dex] & [dex] & [dex]\\
         \hline   
         \ion{Li}{i}  & 0.08 & 0.02 & 0.01 & 0.01\\
         \ion{C}{i}   & 0.01 & 0.02 & 0.01 & 0.00\\ 
         \ion{O}{i}   & 0.03 & 0.04 & 0.00 & 0.01\\
         \ion{Na}{i}  & 0.04 & 0.02 & 0.01 & 0.00\\
         \ion{Mg}{i}  & 0.02 & 0.01 & 0.00 & 0.00\\
         \ion{Al}{i}  & 0.01 & 0.01 & 0.00 & 0.00\\
         \ion{Si}{i}   & 0.03 & 0.01 & 0.00 & 0.00\\
         \ion{S}{i}   & 0.01 & 0.01 & 0.01 & 0.00\\
         \ion{K}{i}   & 0.04 & 0.01 & 0.02 & 0.00\\
         \ion{Ca}{i}  & 0.03 & 0.02 & 0.01 & 0.00\\
         \ion{Ca}{ii} & 0.02 & 0.02 & 0.01 & 0.00\\
         \ion{Sc}{ii} & 0.01 & 0.02 & 0.04 & 0.00\\
         \ion{Ti}{i}  & 0.04 & 0.01 & 0.03 & 0.00\\
         \ion{Ti}{ii} & 0.03 & 0.02 & 0.03 & 0.00\\
         \ion{V}{i}   & 0.08 & 0.01 & 0.01 & 0.01\\
         \ion{V}{ii}  & 0.01 & 0.02 & 0.03 & 0.00\\
         \ion{Cr}{i}  & 0.05 & 0.01 & 0.03 & 0.00\\
         \ion{Cr}{ii} & 0.03 & 0.01 & 0.00 & 0.00\\
         \ion{Mn}{i}  & 0.05 & 0.02 & 0.00 & 0.02\\
         \ion{Fe}{i}  & 0.04 & 0.01 & 0.01 & 0.00\\
         \ion{Fe}{ii} & 0.04 & 0.03 & 0.01 & 0.00\\
         \ion{Co}{i}  & 0.05 & 0.01 & 0.02 & 0.00\\
         \ion{Ni}{i}  & 0.05 & 0.01 & 0.01 & 0.01\\
         \ion{Sr}{ii} & 0.06 & 0.01 & 0.03 & 0.01\\
         \ion{Y}{ii}  & 0.05 & 0.03 & 0.01 & 0.01\\
         \ion{Zr}{ii} & 0.06 & 0.01 & 0.01 & 0.01\\
         \ion{Ba}{ii} & 0.03 & 0.01 & 0.02 & 0.00\\
         \ion{Nd}{ii} & 0.04 & 0.01 & 0.03 & 0.00\\
         \hline \hline
    \end{tabular}
    \label{tab:param_err_abu}
\end{table}

\end{appendix}

\end{document}